\documentclass[twocolumn]{aastex7}
\usepackage[utf8]{inputenc}
\usepackage{hyperref}
\usepackage{tikz}
\usepackage{orcidlink}

\usepackage{natbib,graphicx,amsmath,amsthm,ulem,color,wasysym}
\usepackage{graphicx}	
\usepackage{amsmath}	
\usepackage{amssymb}	
\usepackage{hyperref,float}
\usepackage{cases}

\defcitealias{CG97}{CG97}

\newcommand{\y}{\color{red}}
\newcommand{\z}{\color{orange}}
\newcommand{\w}{\color{blue}}

\newcommand{\be}{\begin{eqnarray}}
\newcommand{\ee}{\end{eqnarray}}
\newcommand{\beqn}{\begin{eqnarray}}
\newcommand{\eeqn}{\end{eqnarray}}
\newcommand{\bi}{\begin{itemize}}
\newcommand{\ei}{\end{itemize}}

\def\g{\, \rm g}
\def\K{\, \rm K}

\def\s{\, \rm s}

\def\cm{\, \rm cm}
\def\m{\, \rm m}

\usepackage{ulem}
\usepackage{xcolor}
\newcommand{\x}{\sout}
\newcommand{\q}{\color{olive}}
\newcommand{\e}{\color{violet}}

\defcitealias{Doi2021}{DK21}

\begin{document}

\title{Geometry of dust rings in protoplanetary disks: the case of LkCa 15}

\author[0000-0002-6794-7480]{Yansong Qian}
\email{yansong.qian@mail.utoronto.ca}
\affiliation{Department of Astronomy \& Astrophysics, University of  Toronto, Toronto, Canada}

\author[0000-0003-0511-0893]{Yanqin Wu}
\affiliation{Department of Astronomy \& Astrophysics, University of 
Toronto, Toronto, Canada}
\email{wu@astro.utoronto.ca}

\begin{abstract}
Dust properties in proto-planetary disks shape the pathways for planet formation. 
Here, we present a method to measure these properties in moderately inclined dust rings. Our method exploits the simple geometric fact that, for such a ring, its ansae appear brighter because our line of sight traverses a longer path through the ring material, and appear broader because the minor axis are foreshortened by projection. 
The resultant patterns of apparent brightness and width can used to constrain three parameters: the intrinsic ring width, its vertical thickness and its optical depth. 
We apply this method to ALMA archival images of the LkCa 15 disk, in Bands 7, 6 and 3. We find that the optical depth of its main ring drops from $1.6$ at 0.89mm to $0.4$ at 3mm. Simultaneously, both the ring width and the ring height decrease from about two to one gas scale heights.
Such wavelength-dependent morphology can only be explained by the presence of multiple grain populations.
If we adopt a simple two-size model, we infer that the ring contains a massive population of small grains (size $\lesssim 20\mu$m; total mass $\sim 100 M_\oplus$) that are broadly distributed, and a less massive population of large grains (size $\gtrsim 200\mu$m) that are more spatially concentrated. 
This large surplus of small grains is not predicted by models of dust coagulation, but it naturally explains the fluffy ring in LkCa 15, and possibly rings in other disks.
\end{abstract}

\section{Introduction}

Over the past decade, high resolution observations have revealed that proto-planetary disks are not smooth, but exhibit a variety of substructures  \citep[e.g.,][]{Marel2013,ALMA2015,AndrewsDSHARP,Long2018}. Among these, gap-and-ring pairs are common and are thought to be the results of planet-disk interactions \citep[e.g.,][]{Zhang2018}, though other explanations exist. 

Near a ring, dust grains tend to migrate towards its center (pressure maximum) and to settle towards the mid-plane. The radial and vertical extents of the dust are therefore good indicators for both the grain sizes and the strength of local stirring,  
important ingredients in establishing the planet-forming potential of disks.

Determining these spatial spreads have motivated a large number of studies, with the vertical extent getting substantially more attention \citep[see][for a recent review]{Villenave2025review}.
For instance, in the iconic HL Tau system, the clear dark gaps sandwiched between bright rings allow \citet{Pinte2016} to infer that the rings are vertically thin, perhaps as thin as $1$au at a distance of 100au. Another example is the pioneering work of \citeauthor{Doi2021}(2021, hereafter \citetalias{Doi2021}), where they demonstrated how, in an inclined ring, variations in the line-of-sight optical depth can be used to deduce its vertical  thickness. Applying to the dust ring of  HD 163296, they showed that the inner  67au ring is vertically as extended  as the gas, while the outer 100 au ring is more settled, with a height smaller than $10\%$ of the gas scale height. 
Their work has spurred a number of  studies \citep{Pizzati2023,Villenave2025,Jiang2025,Martinien2026,Antilen2026}. 
And just like in the case of HD 163296, these studies return a diverse range for the dust scale heights. A common interpretation for such a diversity is that these rings experience (vastly) different strengths of turbulent stirring.

In this work, we are motivated to re-examine the \citetalias{Doi2021} approach, for a number of reasons. 

A necessary assumption in their work is that the ring is optically thin (vertically), or else the optical depth variation would not translate into brightness asymmetry. However, multiple lines of evidence indicate that dust disks are likely optically thick at ALMA bands, especially at Bands 6/7 where most of these studies are carried out. For instance,  disks in multiple star-forming regions
exhibit the so-called size-luminosity relation \citep[e.g.][]{Tripathi2017,Barenfeld2017,Andrews2018,Otter2021}, where disk millimeter luminosities are observed to scale approximately quadratically with their sizes. This indicates that the emission is governed primarily by the emitting surface area and is therefore likely optically thick.
In addition, spectral indices measured in millimeter bands (between 1 to 3mm)
lie close to the blackbody value  \citep[e.g.][]{Pinilla2014,Tazzari2021,Otter2021,Qian2025}. This consideration motivate us to  extend the original \citetalias{Doi2021} work to the optically thick limit. 

Another reason to re-visit this problem is the recognition that the radial width of a ring can also be tapped for information. There are two aspects to this. 
First, the intrinsic radial width of the dust ring provides an additional diagnostic on grain concentration.
Yet it has only been utilized in a handful of studies, i.e \citet{Dullemond2018} in Band 6 and \citet{Doi2023,Sierra2025} in multi-wavelengths. Second, as we show in this work, the apparent radial width of a ring also carries information on the vertical dust height. 
This latter feature has not been formally discussed, but it is implicitly utilized when considering the shape of a gap bordered by two nested rings \citep[see, e.g.][]{Pinte2016,Villenave2022}.
Here, we explicitly include this observable in our model, and show that it breaks model degeneracy and is useful even when only one ring is present.

A third reason to revisit the \citetalias{Doi2021} model is to emphasize the role of beam shape. This has been improperly neglected in the past. Lastly, we will employ our model on the multi-band observations of LkCa 15. We show that this multi-wavelength approach provides unique new insights. We find an alternative explanation for the diverse dust scale heights previously reported.

In \S \ref{sec:method}, we construct a geometric toy-model to connect dust geometry to the observables.  In \S \ref{sec:apply}, we apply the model to the LkCa 15 disk, and present a physical model for the measurements.  We discuss physical implications and caveats in \S \ref{sec:discussion}, before concluding in   \S \ref{sec:conclusion}.

\begin{figure*}
    \centering
    \includegraphics[width=0.85\linewidth]{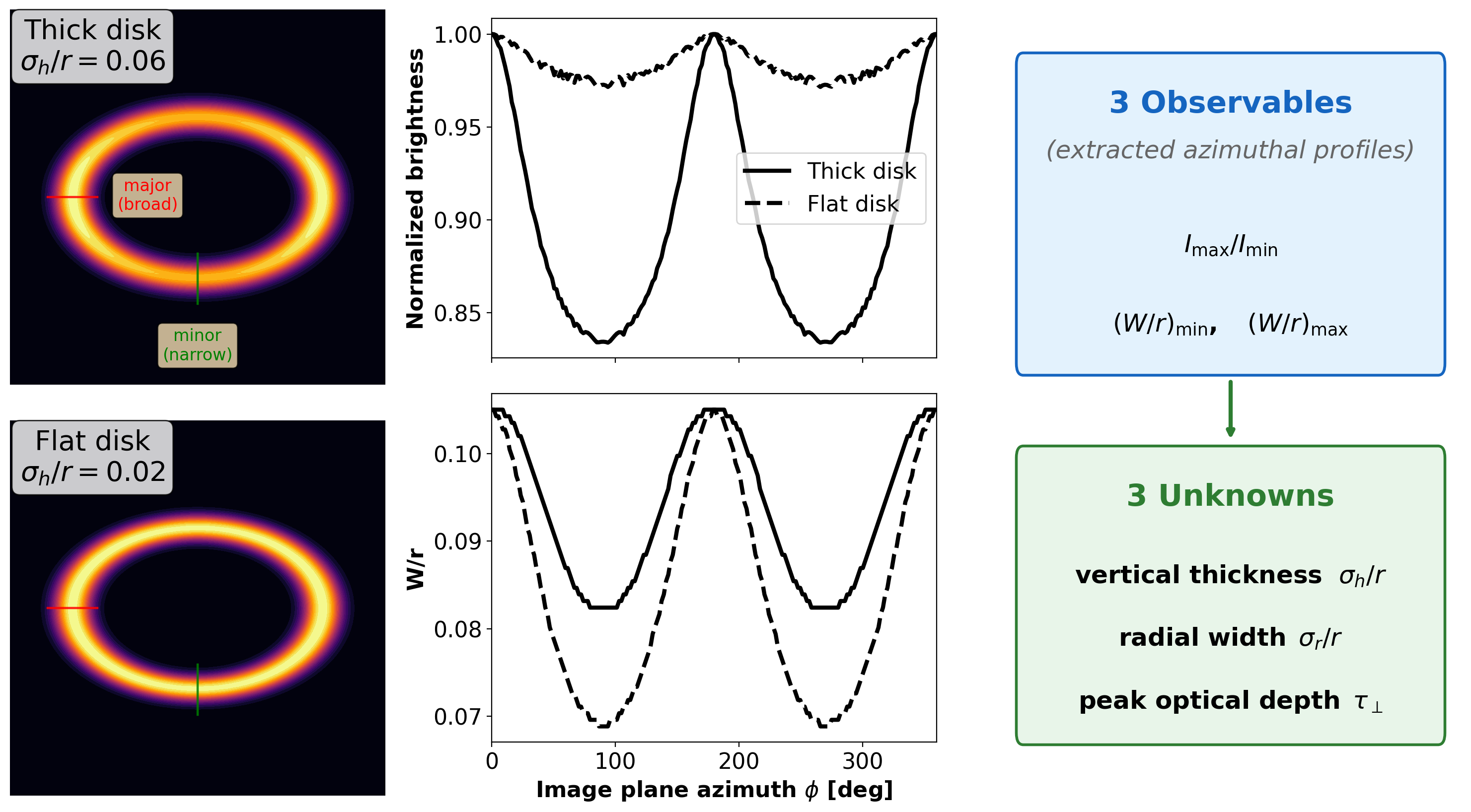}
    \caption{A cartoon to illustrate our geometrical model. Two inclined ($i=50^\circ$) Gaussian rings with similar optical depths ($\tau_\perp = 0.1$) and radial widths ($\sigma_r/r = 0.1$), but different vertical heights, can be distinguished by their azimuthal patterns in apparent brightness (middle upper panel) and apparent width (middle lower panel). Beam effects are not included here. The flatter ring appears more uniform in brightness \citepalias{Doi2021} and has a more fore-shortened minor axis. Including the apparent radial width allows us to infer all three parameters of a Gaussian ring (right panel). A further observable, the absolute flux, yields disk temperature.}
    \label{fig:cartoon}
\end{figure*}

\section{A Toy Ring}
\label{sec:method}

In this section, we follow \citetalias{Doi2021} to construct a geometrical model for an inclined ring. The focus is on how projection effects can produce variations in both its width and its brightness, and how these can be used  to constrain the ring's properties. 

We consider an axisymmetric ring with a dust density that is normal distributed in both the radial and the vertical directions:\footnote{For the radial direction, we note that an actual ring may not be Gaussian in shape, and its inner and outer edges may be asymmetric. Our assumption needs to be refined for well resolved rings.
}
\begin{equation}
    \rho(x,y,z)=\frac{\Sigma_0}{\sqrt{2\pi}\sigma_h }\exp\left[ -\frac{(r-R)^2}{2\sigma_{r}^2}-\frac{z^2}{2\sigma_{h}^2}\right] \,,
    \label{eq:density}
\end{equation}
 where $z$ points at the disk normal, and $r=\sqrt{x^2+y^2}$. Here, $\Sigma_0$ is the two-sided surface density measured at the ring centroid, $r = R$. It is related to the peak (two-sided) optical depth $\tau_{\perp}$ as $\tau_{\perp} = \kappa_d \,\Sigma_0 $, where $\kappa_d$ is the dust absorption opacity  and is assumed to be constant across the ring (for a given population of dust grains). Effects of grain scattering is considered in Appendix \ref{sec:scatter}.
We further assume a constant temperature $T$.
These assumptions allow us to calculate the intensity map of an inclined ring trivially,
\begin{equation}
    I(x_o,y_o)=B_{\nu}(T)\left[1-\exp\left(-\kappa_d\int_{-\infty}^\infty \rho(x,y,z) \,  dz_o\right)\right]\, ,
    \label{eq:intensity}
\end{equation}
where the coordinates ($x_o$,$y_o$,$z_o$) are now in the observer's frame. The axis $z_o$ points along the line-of-sight and is inclined by an angle $i$ from $z$. Meanwhile, $x=x_o$, $y=y_o\cos{i}-z_o\sin{i}$, $z=y_o\sin{i}+z_o\cos{i}$. We then convolve this intensity map (in the observer's frame) with a Gaussian beam to produce the final image.

This final image is to be compared against data. When doing so, instead of working on the 2-D images directly as in some previous works \citep[e.g.][]{Pinte2016,Jiang2025,Antilen2026}, we 
opt for a more transparent and economic method. We will extract only two features from the images, i.e., the ring brightness and radial width as functions of image-plane azimuth. We show that these are sensitive and sufficient to infer the three intrinsic properties of a ring: peak optical depth ($\tau_\perp$), radial width ($\sigma_r/r$) and vertical thickness ($\sigma_h/r$).

A cartoon version of this approach is illustrated in Fig. \ref{fig:cartoon}. 
Fig. \ref{fig:dependency} presents the detailed numerical results for an inclined ring.  To obtain these, we define the azimuthal brightness as the peak surface brightness of a ring at each azimuth angle. At each angle, we also measure the radial width by fitting  a Gaussian form (of dispersion $\sigma_r=W $) to the radial brightness profile. 
In the following, we explain these results in simple words, before applying them to a real disk.

\subsection{Asymmetry in Brightness}
\label{sec:brightness varaition}

\begin{figure}
    \centering
    \includegraphics[width=\linewidth]{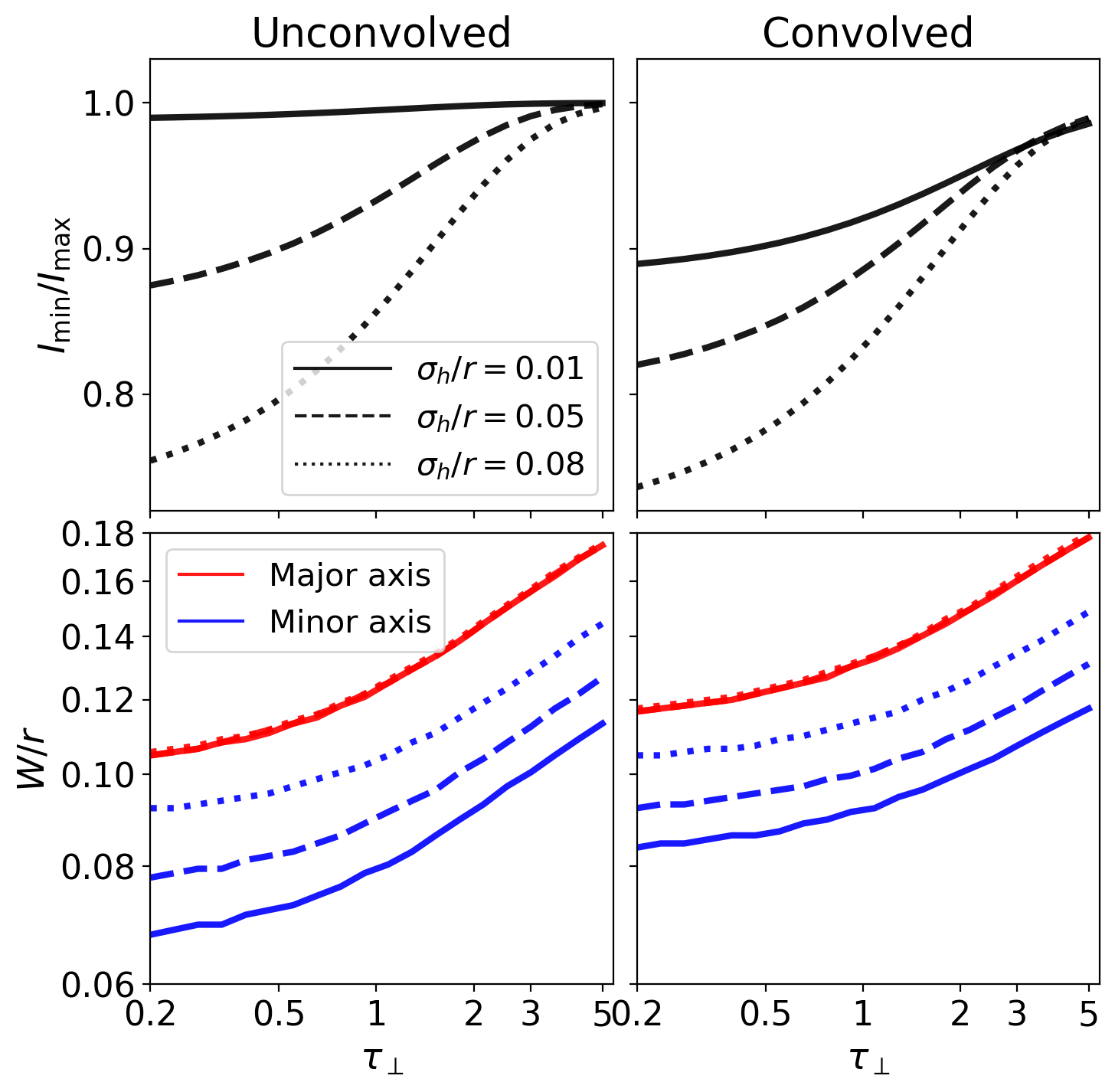}
    \caption{
    Asymmetries in azimuthal brightness (top) and azimuthal radial width (bottom) for our model ring, plotted as functions of ring optical depth, and for different ring heights. All models are viewed at an inclination of $50^\circ$, and have an intrinsic width of $\sigma_{\rm r}/r=0.1$.     
The left panels show results with infinite spatial resolution, while the right ones are convolved with a circular beam of $\sigma_b/r=0.05$. Vertical thickness and beam tend to enhance the brightness contrast, and reduce the width contrast. Increasing optical depth suppresses the brightness contrast and fattens the ring. 
}
    \label{fig:dependency}
\end{figure}

If a ring is optically thin and vertically extended, it appears brighter along its major axis than along its minor axis because the line of sight passes through more material at the ansae. Such an asymmetry is stronger for a ring that is more vertically extended, as was first pointed out by \citetalias{Doi2021}.

For realistic rings, we argue that there are two necessary corrections. 
The first is that ring optical depths are likely close to or above unity, especially in shorter wavelength ALMA bands (see Introduction). The brightness asymmetry should be diminished in such a ring, as the radiation intensity approaches the local blackbody and saturates (see Fig. \ref{fig:dependency}).
The other correction is the effect of an observing beam. Convolution with a wide beam dilutes the brightness of a ring. It is of no concern if the ring has a uniform shape across all angles. However, as we argue below, the ring is typically narrower near the minor axis. Beam convolution dilutes the surface brightness there more severely,  giving rise to an additional brightness asymmetry (see Fig. \ref{fig:dependency}). This effect holds even if the observing beam is round.
And it is significant as long as the beam is not much smaller than half of the intrinsic ring width -- a common happenstance among ALMA observations. 

As a result of these two complications, the brightness contrast now depends not only on the ring height, but also on the ring optical depth and the observing beam. Measuring this contrast alone is no longer sufficient. A new observable is required.

\subsection{Asymmetry in Radial Width}
\label{sec:width variation}

The apparent width of a ring provides this additional constraint. This was little discussed in the past,
possibly because measuring width requires a high spatial resolution \citep[but see][]{Villenave2020,Villenave2025}.

An inclined ring will appear narrower in its minor axes due to fore-shortening. In the limit that the ring is razor-thin ($\sigma_h = 0$), the projected radial width is approximately
\begin{equation} 
W(\phi) \sim \sigma_r \sqrt{\cos^2\phi+\sin^2\phi\cos^2i}\,,
\label{eq:width}
\end{equation}
where $\sigma_r$ is the intrinsic width, $i$ is the inclination angle and $\phi$ the azimuthal angle measured from a major axes in the image plane. So a thin ring that is inclined by $50^\circ$ will exhibit a $\sim 40\%$ width variation from the major to the minor axis.

This width contrast is sensitive to the ring thickness ($\sigma_h$). The thickness broadens the ring everywhere, but affects the minor axis more both because they are already thinner to start with, and because the vertical extension there is parallel to the radial extension. 

Two additional factors -- optical depth and beam convolution -- also affect the apparent ring width but in a different way. 
As shown in Fig.~\ref{fig:dependency}, increasing the optical depth beyond unity 
broadens the apparent widths of a ring, 
because the emission profile is transformed from a Gaussian shape to a top-hat shape.  However, this broadening affects the major and minor axes nearly equally, leaving their width contrast largely unchanged. Beam convolution likewise broadens the ring in a similar manner along both axes. 
These dependencies are all included in our model.

\bigskip

Including the azimuthal dependence of the ring width therefore breaks the degeneracy inherent in the \citetalias{Doi2021} model. As Fig. \ref{fig:cartoon} indicates, we now have three measurements, brightness contrast ($I_{\rm max}/I_{\rm min})$, maximum width ($(W/r)_{\rm max}$), minimum width ($(W/r)_{\rm min}$), to constrain the three intrinsic variables: $\tau_\perp$, $\sigma_r$, $\sigma_h$. The degeneracy is broken. In the following, we will apply our toy model to the disk of LkCa 15.

\section{Application to LkCa 15}
\label{sec:apply}

LkCa 15 is a young T-Tauri star in the Taurus star-forming region, at a distance of 159 pc \citep{Gaia2018}. The star mass $M_*=1.2M_\odot$ \citep{Manara2014} and is likely a couple Myrs old.
It hosts 
two bright rings that are inclined at $i = 50.2^\circ$\citep{Facchini2020}. 
The disk is bright both in scattered light \citep[e.g.][]{Swastik2026} and in thermal emission, with an integrated thermal luminosity of $\sim 0.28 L_*$ \citep{Sierra2025}.
We choose this disk for its large size, and for its extensive coverage at high-resolution in multiple ALMA wavelengths \citep{Long2022, Sierra2025}. These  images allow us to accurately quantify ring asymmetries. Here, we will focus on the prominent inner ring at 69au.\footnote{We also investigate the outer 101au ring in Appendix \ref{sec: outerring}.}
We will first extract ring width and brightness profiles from ALMA images (\S \ref{sec:observation}). We will then present results for dust geometries in \S \ref{sec:res geometry} and construct a simple model to interpret these results (\S \ref{subsec:twopop}). Multi-band observations are instrumental in achieving these interpretations.

\subsection{Observations}
\label{sec:observation}

\begin{table*}[]
    \centering
    \begin{tabular}{cc|cccc|cccc}
    \hline
    \hline
    \multicolumn{2}{c|}{Observation} & \multicolumn{4}{c|}{Multi-wavelength Joint Fit} & \multicolumn{4}{c}{Single-Band Fit}\\  
Band &$\lambda_{obs}$ & $T_{\rm ring}$ & $\tau_\perp$ & $\sigma_r/r$ & $\sigma_h/r$ &$T_{\rm ring}$ & $\tau_\perp$ & $\sigma_r/r$ & $\sigma_h/r$ \\
\hline

7 & 0.88mm & $13.0^{+ 0.6 }_{- 0.6 }\K$ &$1.6^{+ 0.3 }_{- 0.3 }$ &$0.121^{+ 0.008 }_{- 0.007 }$ &$0.080^{+ 0.007 }_{- 0.007 }$ & $13.0^{+0.7}_{-0.4}\K$ & $1.5^{+0.3}_{-0.3}$ & $0.123^{+0.009}_{-0.007}$ & $0.079^{+0.007}_{-0.008}$\\
& &&&& &&&&\\
6 & 1.34mm & (same)
&$1.1^{+ 0.3 }_{- 0.2 }$ &$0.092^{+ 0.005 }_{- 0.006 }$ &$0.063^{+ 0.007 }_{- 0.008 }$ & $14^{+2}_{-1}\K$ & $0.9^{+0.4}_{-0.3}$ & $0.096^{+0.007}_{-0.008}$ & $0.061^{+0.007}_{-0.008}$\\
& &&&& &&&&\\
3 & 3.08mm & (same)
&$0.35^{+ 0.03 }_{- 0.03 }$ &$0.078^{+ 0.004 }_{- 0.004 }$ & $0.059^{+ 0.006 }_{- 0.006 }$ &$15^{+7}_{-4}\K$ & $0.3^{+0.2}_{-0.1}$ & $0.078^{+0.005}_{-0.005}$ & $0.059^{+0.006}_{-0.006}$\\
\hline   
\hline
    \end{tabular}
    \caption{Best-fit parameters for the geometric model, with 1$\sigma$ uncertainties, and in different ALMA bands, for the inner (69au) ring of LkCa 15.  For comparison, the vertical gas height at the ring center is $h/r \sim 0.06$. The observing beam is $\sigma_b/r=0.059$ for all three bands. }
    \label{tab:fitting}
\end{table*}

We use the publicly available Band 3 (97.5 GHz, 3.08 mm), Band 6 (224 GHz, 1.34 mm), and Band 7 (340 GHz, 0.88 mm) images from \citet{Sierra2025}. The Band 6 and 7 data were originally reported by \citet{Long2022}. 
For consistency across wavelengths, \citet{Sierra2025} have convolved all images to a common circular beam with a FWHM of 60mas FWHM.

\begin{figure*}
    \centering
    \includegraphics[width=0.9\linewidth]{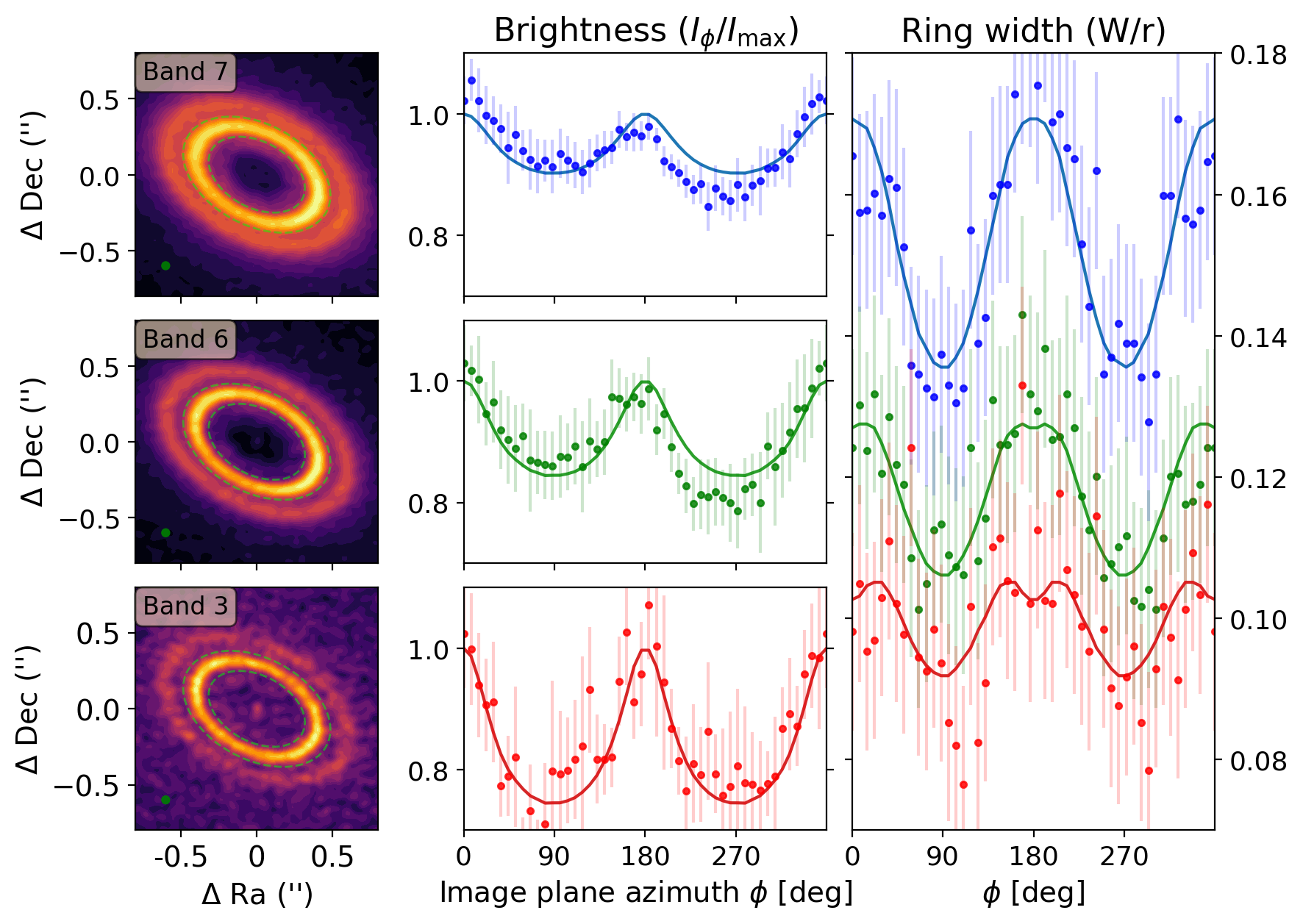}
    \caption{Images and corresponding azimuthal profiles (brightness in the middle and radial width at the right) for the 69au ring in three ALMA Bands. The dashed curves in the left images enclose regions we use to extract the azimuthal profiles of the 69au ring. Uncertainties on brightness only include the rms noise, while uncertainties on width are taken to be 1/3 of the beam size. The corresponding best-fit models are shown as colored curves, based on the joint fit from Table \ref{tab:fitting}. Each brightness profile is normalized by its maximum intensity. Band 3, compared to other bands, requires a narrower/thinner Gaussian ring that is more optically thin. 
}
    \label{fig:phiobs}
\end{figure*}

We measure azimuthal brightness and radial width the same way as for our toy ring (see \S \ref{sec:method}).
To account for correlation among data points in the image, we have adopted a spacing in the azimuthal direction of $1\times$FWHM of the beam. In the radial direction, however, we adopt three grid points per beam FWHM to adequately resolve the radial brightness profile and measure the ring width. 
At each azimuth angle, we first find the peak brightness, $I_\phi$. We then extract the radial profile inside a region that is falls $\pm 1.5\times$ beam FWHM of the ring peak (see Fig. \ref{fig:phiobs}). Restricting the fitting region in this way reduces contamination from adjacent rings. We then fit the resulting radial profile using a single Gaussian and obtain its dispersion $W$. The uncertainty in brightness is estimated as the standard deviation of the pixel values within a synthesized beam centered on the peak. 
For simplicity, we adopt the uncertainty on the ring width to be $1/3$ of the beam size.

Our resultant measurements are shown in Fig.~\ref{fig:phiobs}. There are clear azimuthal variations in both the ring width ($W$) and the ring brightness ($I_\phi$), by amounts that significantly exceed the measurement uncertainties. The minor axes are always dimmer and narrower, as expected. Moreover, the brightness asymmetry is larger in longer wavelengths, whereas the ring looks broader in shorter wavelengths.  It is clear that a single set of parameters cannot explain all bands.

To proceed, we first assume that the ring in each band has a unique set of parameters ($T_{\rm ring}$, $\tau_\perp$, $\sigma_r$ and $\sigma_h$). 
We define the log-likelihood
\begin{equation}
    \log L=-\frac{1}{2}\sum\frac{(W_{obs}-W_{model})^2}{E_{W}^2}+\frac{(I_{obs}- I_{model})^2}{E_{I}^2}\,,
    \label{eq:mcmc}
\end{equation}
where $E_W$ and $E_I$ denote the corresponding observational uncertainties. We then perform Markov Chain Monte Carlo (MCMC) sampling, using the python package \texttt{emcee} \citep{emcee} to find the maximum likelihood. Each chain has 64 walkers and are run for 5000 steps. The first 1000 steps are discarded as initial burn-in. 

Results for individual bands are listed in Table \ref{tab:fitting} (the `Single-Band Fit'), where we also infer ring temperature using the observed fluxes.\footnote{These do not yet include calibration errors. See the `Joint Fit' below.} 
As expected, we find that ring properties vary with the observing wavelength: while the ring temperature is largely constant across all bands, its intrinsic width, height and peak optical depths all decrease with increasing wavelength. 

This inspire us to perform a joint analysis, insisting a single temperature across all bands (the `Multi-wavelength Joint Fit' in Table \ref{tab:fitting}). To account for calibration errors ($\sim 10\%$ in Bands 6/7, and  $\sim 5\%$ in Band 3, ALMA Technical Handbook), we introduce a calibration factor $c$ into the likelihood function by replacing $I_\mathrm{model}$ with $c*I_\mathrm{model}$ in Eq. \ref{eq:mcmc} for Bands 6/7, but ignore the smaller error on Band 3. We impose a Gaussian prior on the two calibration factors, $c\sim \mathcal{N}(1,0.1)$. The total log-likelihood is then obtained by summing the likelihoods from all three bands. The MCMC corner-plot for this joint fitting is shown in the appendix (Fig. \ref{fig:corner}). 

In the Joint Fit, we find that all ring parameters remain largely the same as in the single-band fit, but with slightly improved uncertainties. We stress that our model is purely geometrical -- there is no assumption on the opacity law or the underlying temperature. All information is contained in the apparent shape of the ring.

\subsection{Results on dust geometry}
\label{sec:res geometry}

Before turning to a physical interpretation, we first briefly comment on our results and contrast them with existing measurements in the literature. 

We compute the expected gas scale height ($h$) at ring center. Under hydrostatic equilibrium, this is given by
\begin{equation}
    \frac{h}{r}=\frac{c_s}{v_{K}}=\left(\frac{k_b}{\mu m_HGM_*}\right)^{1/2}\sqrt{rT}\,,
\end{equation}
which yields $h/r \approx 0.06$ using the parameters of LkCa 15 ($M_* =1.2,M_\odot$, $r=69$ au, $\mu=2.3$, and $T=13$ K).

The measured radial widths are generally larger than this, dropping from about $2h$ to $1h$ from Band 7 to Band 3. Our values agree well with those obtained in \citet{Sierra2025} using the Frankenstein reduction (fractional widths $0.115$, $0.095$, $0.062$), but fall below those using the CLEAN pipeline as the latter include beam convolution.

Our measured heights ($\sigma_h$) are roughly 30\% below the radial widths. At Bands 6 and 3, these heights are comparable to $h$, while sitting above $h$ at Band 7. This gives the impression that all dust grains are well mixed with the gas. However, we argue later that such an inference may be incorrect. 

At Band 6, two previous studies have also reported dust height. Our value ($\sigma_h/r = 0.063^{+0.007}_{-0.008}$) is consistent with the lower limit  ($\geq 0.067$) reported by \citet{Villenave2025}, but inconsistent with the value of $\sigma_h/r \sim 0.022$ reported by \citet{Jiang2025}. We explain how these differences may arise in \S \ref{sec:discussion}.

\begin{figure}
    \centering
\includegraphics[width=\linewidth]{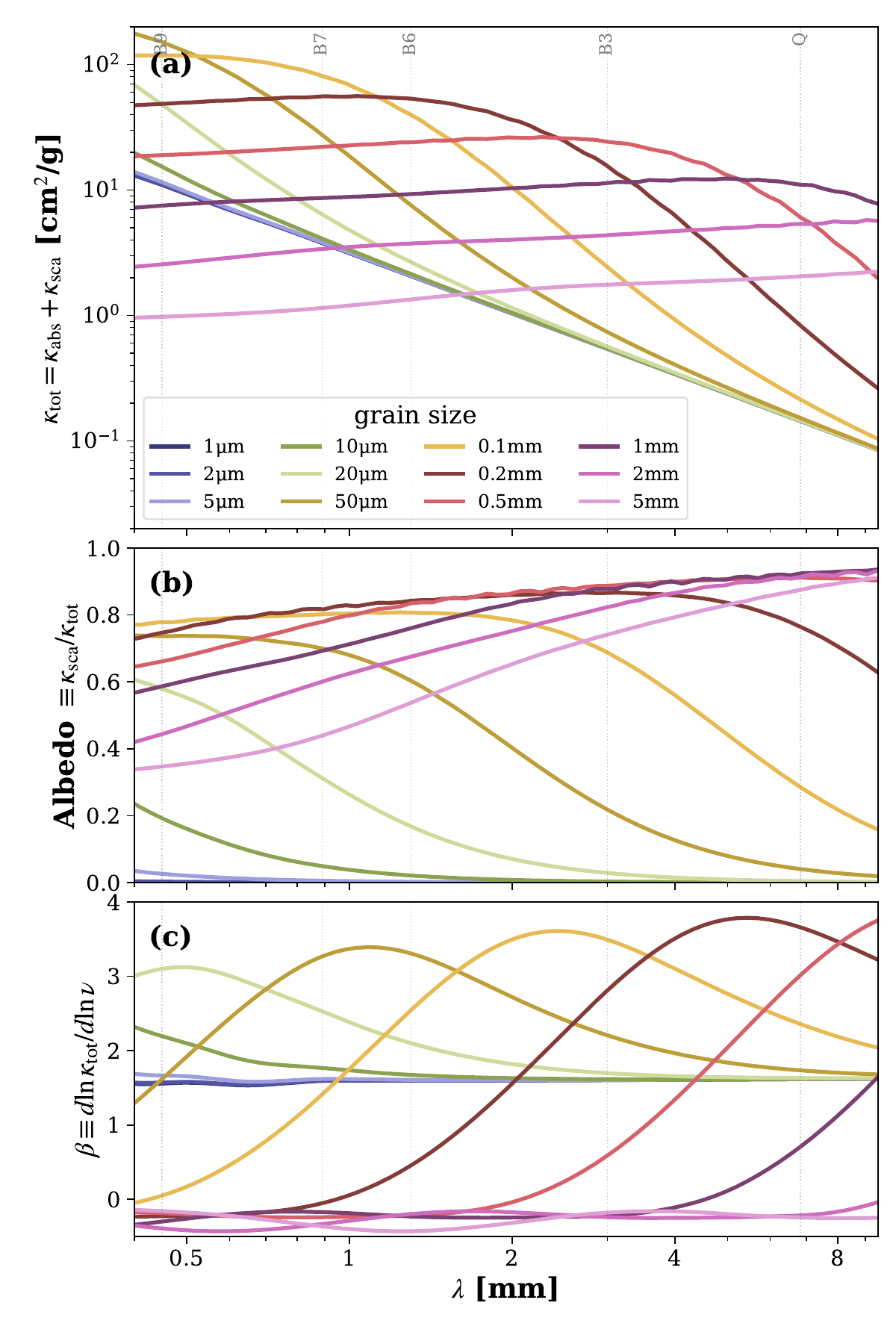}
    \caption{Optical properties for grains of different sizes. We adopt the DIANA composition and evaluate for a narrow Gaussian distribution around the said size (see text). In ALMA Bands 7/6/3, all grains below $\sim 20\mu m$ have similar opacities, while larger grains are strong scatterers.
    }
    \label{fig:diana}
\end{figure}

Lastly, as an improvement over previous studies, our approach also allows us to measure the values of the peak optical depths. 
We find them to also decreases with increasing wavelength, transitioning from moderately optically thick to optically thin across the three bands.  This finding is consistent with the spectral indexes reported by \citet{Sierra2025}, who found that $\alpha \lesssim 2$ between Bands 6 and 7, while $\alpha$ rises to $\sim 2.3$ between Bands 6 and 3. Moreover,  the obtained optical depths scale with wavelength approximately as $\tau_\perp \propto \lambda^{-1.2}$. Such an opacity index is not very different from that expected of small grains ($a \ll \lambda$): $\beta \approx 1.5$ (DIANA grains, see Fig. \ref{fig:diana}), hinting already at the role of small grains. In fact, without the information on the varying ring width/height, one would have easily believed that small grains are all there is.

\bigskip 

In summary, we find the 69au ring to be broad and thick, with 
 a number of intriguing variations. In the following, we propose an explanation for these observations.

\subsection{A two-size Model}
\label{subsec:twopop}

If the ALMA emission is dominated by a single grain population, our measured $\tau_\perp(\lambda)$ would directly yield the grain size. However, the fact that the ring is seen to vary in width and height across wavelength suggests that this assumption cannot hold. Instead, our observation demands that there must be multiple dust populations, each with its own spatial distribution. 

Here, we build the simplest model, one where there are exactly two populations of grains. We further assume each population has a single size. We call this the `two-size model', with sizes $a_{\rm small}$ and $a_{\rm big}$.
Each dust population has densities $\rho_{\rm small}(r,z)$ and $\rho_{\rm big}(r,z)$ that are distributed according to equation (\ref{eq:density}). These introduce 6 unknowns: peak surface densities $\Sigma_{\rm small}$, $\Sigma_{\rm big}$; fractional radial widths $(\sigma_r/r)_s$, $(\sigma_r/r)_b$; and fractional vertical heights $(\sigma_h/r)_s$, $(\sigma_h/r)_b$.
The `small grains' likely dominate in the shorter wavelengths and are responsible for the fluffy look of the ring; while the `big grains' dominate in the longer wavelengths and are more narrowly distributed. 

Our simplifications are necessitated by the small number of constraints: a total of 9 measurements ($\tau_\perp$, $\sigma_r$ and $\sigma_h$ in 3 ALMA Bands). A two-size model, at the same time, already demands 8 parameters. 
Moreover, our single-size assumption differs from the more common approach of a power-law size distribution. We argue that ours is more physical, as one expects the spatial distribution to depend on grain size, or in other words, different sized grains should be spatially segregated.

\begin{figure*}[t]
    \centering
\includegraphics[width=\linewidth]{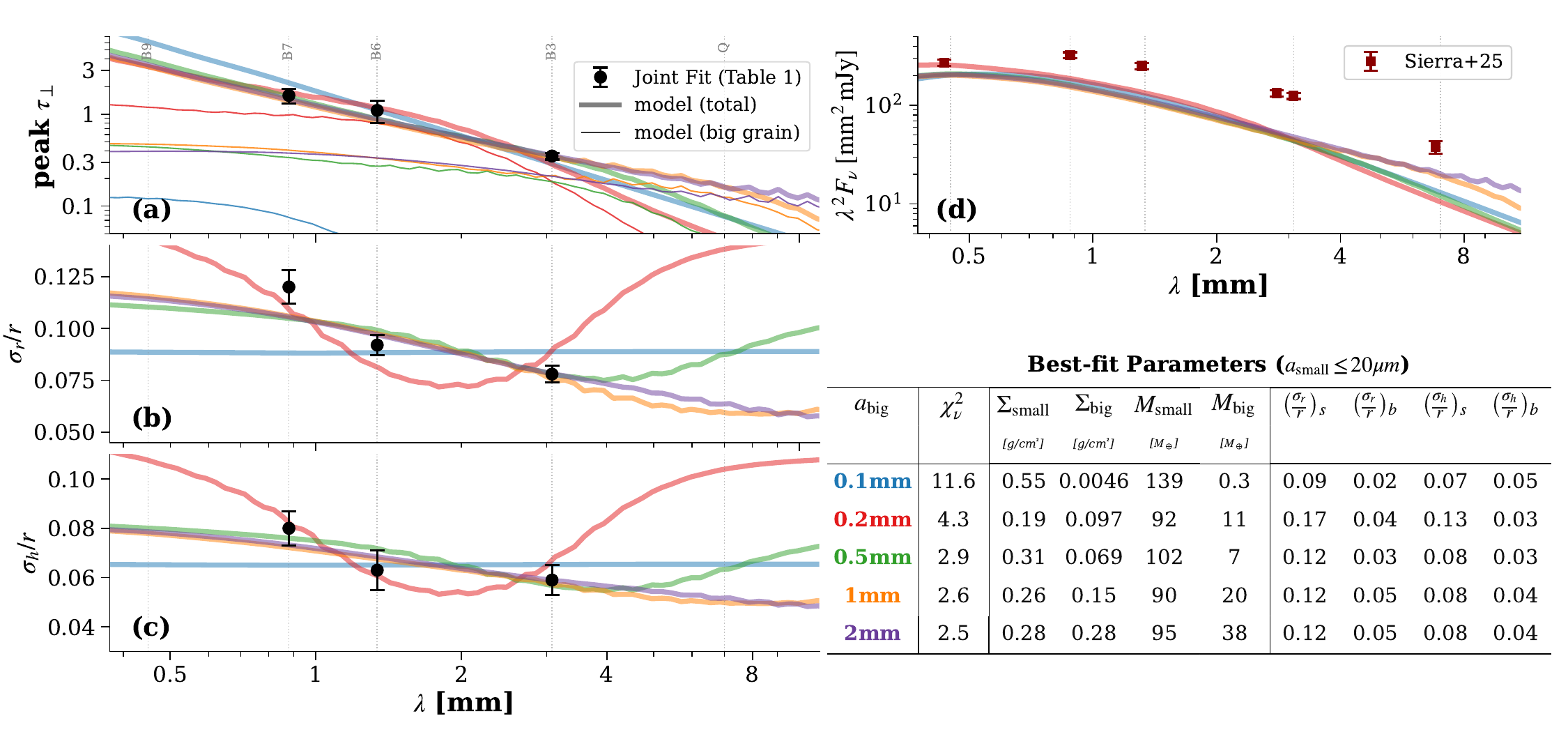}
    \caption{Explaining our measurements (black circles) using a two-size dust model. At each size of large grains ($a_{\rm big}$, colored as in the embedded table), the best-fit model is drawn as a function of wavelength. Panels a) peak optical depth, with the total value in thick curves, and the big grain contributions in thin curves; b) effective radial width; c) effective vertical height; d) resultant SED, calculated for a face-on ring. For the last, we have also over-plotted the observed full-disk values taken from \citep{Sierra2025}. 
    The peak surface densities in the table are two-sided, and we obtain the total grain masses by integrating these over the respective ring widths.
    }
    \label{fig:fitting}
\end{figure*}

A few more choices in our model construction are listed here.

\begin{itemize}  

\item Grain sizes. We choose our small grains to be $a_{\rm small} = 1\mu m$. The exact size, as long as $a_{\rm small} \leq 20\mu m$, is irrelevant,
thanks to the convenient property that all such grains have nearly identical millimeter opacities (see Fig. \ref{fig:diana}).\footnote{In fact, the `small grains' can be an arbitrary combination of these grains (e.g., power-laws) without affecting the results.} The abundant existence of such small grains has been inferred from scattered light studies (see below). We are also prompted to adopt such small grains by the observed opacity index of the observed ring ($\tau_\perp \propto \lambda^{-1.3}$).
Sizes for the big grains ($a_{\rm big}$), on the other hand, is a quantity that we will explore.

\item Grain opacity. We adopt the DIANA composition \citep{Woitke2016} for our dust grains. Other compositions do not qualitatively change our conclusions.\footnote{Switching to the DSHARP composition \citep{Birnstiel2018}, for instance, raises the mass budget by about a factor of $3$.} By the word `single-size', we actually refer to a narrow size-distribution with a Gaussian width of $asig = 0.4$ in the dust opacity code Optool \citep{Optool}. The opacity and opacity index for some relevant sizes are shown in Fig. \ref{fig:diana}. The same figure also exhibits the scattering albedo. These are near unity for large grains in millimeter. 

\end{itemize}

We now obtain properties for the mixture ring.
The peak optical depth is obtained by summing over the two sizes. The effective Gaussian width and height are determined by the best Gaussian fit to the sum of the two Gaussians, weighted by the respective contribution to the local optical depth. This is not strictly appropriate and may fail when the mixture departs significantly from a single Gaussian, e.g., when one of the populations is much narrower than the other.

To match the observed parameters, we define a reduced $\chi_\nu^2$ 
\begin{equation}
    \chi_\nu^2 = {1\over (N_{\rm obs} - N_{\rm var})}\sum_{\rm Bands, i}  
    \frac{\left(X_{\rm model} - X_{\rm obs}\right)^2}{\delta X^2}\, ,
    \label{eq:definechi}
\end{equation}
where the summation is over Bands 7/6/3, and over the variable index $i$, where $i=1,2,3$ correspond to $\tau_\perp$, $(\sigma_r/r)$ and $(\sigma_h/r)$ respectively. The observed values ($X$) and their errors ($\delta X$) are taken from the Joint Fit in Table \ref{tab:fitting}. The degree of freedoms are $N_{\rm obs}= 3 \times 3 = 9$ and $N_{\rm var} = 6$. This exercise is repeated for each choice of $a_{\rm big}$.  We use the tool `differential evolution' in SciPy to search for minima in $\chi^2_\nu$. Often the convergence is straightforward and even brute-force grid search suffices. 

The fitting results are presented in  Fig. \ref{fig:fitting}, where an embedded table shows our best-fit parameters at every choice of $a_{\rm big}$. We summarize our results below and discuss their implications in \S \ref{subsec:thickness}:

\begin{itemize}

\item {\bf small grains:} one of our most interesting finding is that small grains dominate much more than expected. Their total mass is of order $100M_\oplus$,\footnote{This is based on the standard DIANA opacity. More absorbent grains, e.g., with a higher carbon content, will reduce this estimate.} about ten times that in large grains. They also account for most of the  emission in Band 7 and (sometimes) Band 6. These grains are broadly distributed, explaining the large ring width/height that we find in these wavelengths.

Such a large mass in the small grains is surprising (see more below). However, this claim is supported by studies at other wavelengths.
First, scattered light. 
\citet{Swastik2026} carried out the most comprehensive modeling of infrared scattering in LkCa 15 and reported that the scattering grains in the 69au ring are one to ten microns in size (depending on porosity), and have a total mass of $\sim 10 M_\oplus$ if assuming a scale-height $\sigma_h/r = 0.1$. This can be made compatible with our mass estimate, when one adopts a smaller scale-height,\footnote{The LkCa 15 disk reprocesses $\sim 28\%$ of the stellar flux to far infrared. If including also scattering albedo, the scattering surface is likely very tall. This requires a large amount of small grains. A smaller dust scale height exacerbates this problem.} and when one includes grains larger than the scatterers.
Second, the Band 9 ($450\mu m$) image of LkCa 15 \citep{Leemker2022} shows one broad ring that spans both the 69au and the 101au rings in Band 7. This indicates that small grains dominate Band 9 and are shinning brightly even in the gap between the two rings. \footnote{A similarly filled-in Band 9 is also observed in the HL Tau disk \citep{Guerra-Alvarado2024}.} This again suggests a large mass in the small grains. 

We can also compare the above mass estimate to the gas mass. For the whole disk, \citet{Jin2019} previously reported a gas mass of  $100 M_J$, while \citet{Sturm2023} reduced it to $10^{+10}_{-4} M_J$. Taking the latter value and assuming that it is 
all associated with the 69au ring, and that $Z = 0.01$, we expect a total dust mass of $20-60 M_\oplus$, within a factor of two of our estimate. Most of it, we seem to find, is lodged in small grains.

Lastly, we infer a scale-height for small grains that slightly exceeds the gas scale height ($\sim 1.2h$).

\item {\bf big grains:} 
we cannot uniquely determine the value of $a_{\rm big}$. Any model with 
 $a_{\rm big} > 200\mu m$ can reasonably explain our measurements, while grains with a smaller $a_{\rm big}$ fail (large $\chi^2$) because their opacities are too similar to those of the `small grains'.  

We can, however, speak to the big grains' spatial distribution. They have to be concentrated near the ring center ($\sigma_r/r \sim 0.03$) and to the disk mid-plane (a similar $\sigma_h/r$). 
Since the opacity in Band 7 is largely dominated by small grains, while Band 3 by large grains, one observes an increasingly narrow/thin ring at longer wavelengths. This observed change allows us to recover information at scales smaller than the beam width ($\sigma_B/r \sim 0.05$). 

The mass of the big grains lies about ten times below that of the small grains. It increases in models with a larger $a_{\rm big}$, because larger grains are less efficient emitters.

\item {\bf SED:} the spectral-energy-distribution of LkCa 15 has been well characterized \citep[see, e.g.,][]{Sierra2025}. Many SED fits have been attempted \citep[e.g.][]{Sturm2023,Sierra2025,Swastik2026}, 
yielding dust parameters that are drastically different from our solutions here. So to validate our model, we will also compute disk fluxes. We calculate the emission assuming the 69au ring is face-on, so the radiative transfer is only 1-D. 
The results are presented in Fig. \ref{fig:fitting}, and compared against the full-disk integrated SED collected by \citet{Sierra2025}.\footnote{We are forced to do so because we lack the resolved images in Band 9 and Q.} This is not an apple-to-apple comparison: our model ring is assumed to be face-on, while the real disk is inclined and contains multiple rings.
The main message, however, is that the SED is also largely invariant to the choice of $a_{\rm big}$. This should be caution enough for studies that claim to determine grain sizes, based purely on SED.

In that panel, we plot SED in unit of $\lambda^2 F_\nu$. For a blackbody emission in the Rayleigh-Jeans limit, this should be flat with wavelength. We see that emissions in Bands 9/7/6 are indeed largely flat, supporting the claim that these bands are optically thick. In contrast, fluxes in Bands 3/Q fall with wavelengths roughly as $F_\nu \propto \lambda^{-3}$. In our model, this results both from the lowering of the optical depth with wavelength, and from the reduction in emitting area as one sees mostly large grains in these bands.

\end{itemize}

\bigskip

So in summary, we find that a two-size model can be chosen to explain our findings on the 69au ring. The fact that this ring looks broad and fluffy is likely explained by the massive small grains. And the fact that it becomes narrower and thinner at longer wavelength is because big grains increasingly take dominance.

\section{Discussions}
\label{sec:discussion}

Here, we first discuss the measurement of dust scale heights, in order to put our work in context with similar studies. We then move on to discuss the physical implications of our results.

\subsection{Measuring Dust Heights}

A variety of methods have been developed to measure dust heights, and a diversity of results have been obtained \citep[see][for a review]{Villenave2025}. 
We comment on the improvements we make over these past studies. 
\begin{itemize}

\item In this work, we emphasize the importance of measuring the radial width. The radial width is as informative on  turbulence stirring as the vertical height. In addition, the variation of width over azimuth allows one to measure the disk height. 

\item We argue that the beam size matters, even when the beam is round. This contrasts with previous understanding \citep{Doi2021,Villenave2025} where only elliptical beams are thought to affect the brightness asymmetry. This is because the ring width is the smallest at the minor axis (fore-shortened by $\cos i$), so a beam dilutes the flux most there. A larger beam enhances the brightness asymmetry (Fig. \ref{fig:dependency}), and may mislead one into believing that the ring is thicker than it really is.
This effect is more severe for smaller rings (relatively larger value of $\sigma_B/r$). We suspect that it may, at least partially, explain why \citet{Jiang2025,Villenave2025} reported that in systems with two-rings, the inner rings tend to be thicker.

\item We explicitly solve for the optical depth of the ring, without any assumption on the dust temperature. This departs from past studies where optical depth is obtained only by adopting a temperature model. In addition to being model-dependent, the conventional approach also loses its sensitivity when the disk is optically thick.

The high optical depth we find in Band 6 explains why our height value differs from that reported in \citet{Jiang2025}: $\sigma_h/r = 0.063$ as opposed to $0.022$ -- they fixed the optical depth to $\tau_\perp = 0.32$ at Band 6. As Fig.~\ref{fig:dependency} demonstrates, an optically thin ring that is geometrically thin can look as asymmetric in brightness as an optically thick ring that is much puffier. 
Fig.~\ref{fig:dependency} demonstrates that it is impossible to obtain a thickness measurement when the optical depth is not known -- one could only hope for a lower-limit.

\item  Unlike previous studies that use the entire 2-D image \citep[e.g.][]{Villenave2025,Jiang2025,Antilen2026}, we explicitly focus on two quantities: peak brightness and apparent radial width.
In so doing, we channel our attention to quantities that are the most defining. It also allows for a transparent and economic exploration of the parameter space.

Figure 3 of \citet{Villenave2025} also illustrated how the width at the minor-axis can be used to constrain ring thickness, the effect we explicitly investigate here. However, unlike our measurements here, their study of the 69au ring only yields a lower limit to the dust height. This may be related to their having a limited number of models and the best model lies at the edge of their model grid (Villenave, private communication).

\item We emphasize the utility of multi-band observations, following the works of \citet{Doi2023,Sierra2025}. Previously, \citet{Sierra2025} have established that the width of the 69au ring decreases with wavelength and attributed it to differential trapping of dust grains. Here, we further reveal that the dust height also decreases with wavelength. Both lengths drop by a factor of 2 from Band 7 to Band 3. Such a stereo vision allows us to draw physical conclusions that are not possible before, and
leads us to a new interpretation for the fluffy disk (see next section), one that is not related to strong stirring.

In contrast, previous works, including population studies by 
\citet{Pizzati2023,Villenave2025,Jiang2025,Martinien2026,Antilen2026}, predominately use data from only one band (typically Band 6). Collectively, these works found a broad range of $\sigma_h/r$ that spans almost  two decades (from  $10^{-3}$ to $\sim 10\%$), 
an example being HD 163296 with its two  rings differing in thickness by a factor of 10 \citepalias{Doi2021}.
If these heights are related to dust stirring, they would require a dramatic (and puzzling) range in the turbulence strength. Again, we argue below against the stirring interpretation. 

\end{itemize}

It will be useful to apply our method to more disks, especially as high-resolution, multi-band observations become increasingly common.

\subsection{What do our Results Mean?}
\label{subsec:thickness}

Large grains are expected to drift rapidly towards the pressure maximum and to settle towards the disk mid-plane. We calculate the speeds of these movements using a model with  $\Sigma_{\rm gas} = 30\g/\cm^2$ (two-sided) at the ring center ($69$au) and a Gaussian width $w_r/r = 0.12$ (or $w_r/h \sim 2$),  same width as that inferred for the small grains (Fig. \ref{fig:fitting}). The total gas mass in the ring is $20M_J$, consistent with that inferred for LkCa 15 by \citet{Sturm2023}.
We take the gas temperature to be a constant $13\K$.

Defining the Stokes number for a grain of size $a$ as $St \equiv a\times (\rho_{\rm bulk} \Omega_k /\rho_{\rm gas} c_s) $, where $\rho_{\rm bulk}$ is the grain bulk density and $\rho_{\rm gas}$ the local gas density, we obtain that  $St \sim 10^{-5}$ for a $1\mu$m grain at the midplane of the ring center. The Stokes number rises with height as the upper atmosphere is more dilute. 

\begin{figure}
    \centering
    \includegraphics[width=0.9\linewidth]{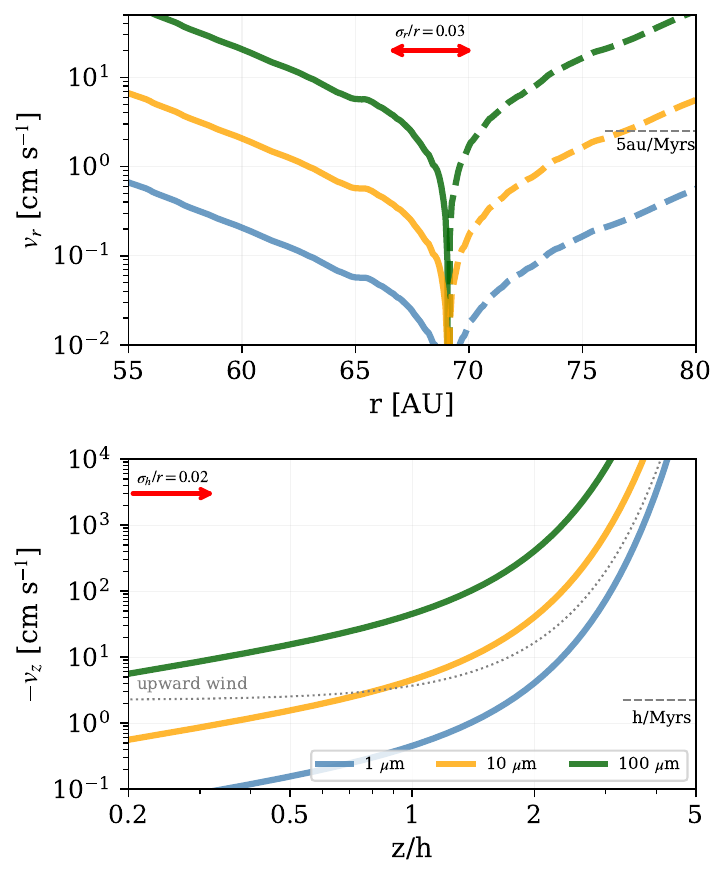}
    \caption{Speeds of radial drift (top panel) and vertical settling (bottom) for three representative grain sizes near the 69au ring. The red arrows indicate our deduced geometry for the big-grains. Micron-sized grains can stay afloat below $z/h = 2$, and to an even larger height if considering lofting by an `upward wind' (see text). Larger grains, in contrast, should have been vertically settled. The situation is similar in the radial direction.
    }
    \label{fig:timescales}
\end{figure}

The characteristic timescales for vertical settling and for radial migration are \citep[see, e.g.][]{Chiang2010}
\begin{eqnarray}
t_{z} & \equiv & {z\over{v_z}} \approx {1 \over{\Omega_k St}} \sim 10^6 {\rm yrs} \, \left({{St}\over{10^{-4}}}\right)^{-1} 
 \nonumber \\
t_{r} & \equiv & {{w_r}\over{v_r}} \sim \left(\frac{h}{r}\right) {1\over{\eta \Omega_k St}} \sim 2\times 10^6 {\rm yrs} \,
\left({{St}\over{10^{-4}}}\right)^{-1} \, ,
\end{eqnarray}
where $\eta \sim (h/r)^2 {{d\ln P}\over{d\ln r}}  \sim (h/r) (h/w_r)$, and we have evaluated with $h/r = 0.06$ and at the limit $St \ll 1$. The corresponding velocities are shown in Fig. \ref{fig:timescales}. 

We first consider the state of the small grains. Assuming a system age of a few Myrs, these estimates suggest that small grains (up to tens of $\mu$m) in the bottom scale-height have neither settled nor migrated. We expect them to show large radial width and vertical thickness, as is indeed found in the two-size model. On the other hand, they should settle rapidly at altitudes above $2h$ where the gas is more dilute \citep[also see][]{Dullemond2004}. Their presence there then requires active stirring or wafting. The latter can be provided by, e.g., an up-ward blowing wind. 
Imagine a steady wind that has a speed of $h$/Myrs near the base (such that all mass is blown away after 1 Myr). It quickens at high altitude as the gas density drops, as $v_z \propto 1/\rho_{\rm gas}(z)$. Fig. \ref{fig:timescales} shows that such a wind can counteract settling and waft small grains up to a large number of scale-heights. An obvious candidate for such a wind is the MHD disk wind \citep[e.g.][]{Bai2013}.

What about the `big grains' ($a_{\rm big} \geq 200\mu m$)? These grains are seen to be more compactly distributed than the gas, with $\sigma_r/r \sim \sigma_h/r \sim 0.02$ (Fig. \ref{fig:fitting}). 
Fig. \ref{fig:timescales} indicates that this requires some form of active maintenance.

If we assume this maintenance is by turbulent stirring, using expressions from \citet{Dullemond2018} for radial stirring, and from \citet{Dubrulle1995,Youdin2007} for vertical stirring, we find a comparable magnitude of stirring strength in both directions,
\begin{equation}
\alpha_r \sim \alpha_z \sim  10^{-1} St \sim 10^{-4} \left({{a_{\rm big}}\over{100 \mu m} }\right)\, .
\label{eq:alpha}
\end{equation}
We are largely ignorant as to the true value of $\alpha$, in particular in dusty rings.
An MHD simulation by \citet{Xu2022}  reported $\alpha \sim 10^{-3}$, but the physical conditions (e.g., the large amount of small grains) in the 69au ring may differ from their set-up.

So overall, the large heights and widths observed in the 69au ring of LkCa 15 is most naturally explained by the fact that ALMA emission is dominated by small grains that are tightly coupled to gas, rather than an anomalously strong turbulent stirring. The large grains are more concentrated and can be stirred by an isotropic turbulence.

\subsection{Puzzles and Caveats}

Our result -- that small grains dominate the dust mass -- poses a new puzzle.

Small grains should  collide at a high rate and grow in size. The mean collision time,
\begin{eqnarray}
t_{\rm mfp} & \approx & { {1}\over{n \sigma \, \Delta V}} 
\sim    \left({{\Sigma_{\rm gas}}\over{\Sigma_{\rm dust}}}\right) {{h St}\over{f \, \Delta V}} \sim {1\over{Z\,f \,\Omega_{\rm k}}}
\nonumber \\
& \sim & {{8000\, {\rm yrs}}}\,\times  f^{-1}  \, \left({Z \over{0.01}}\right)^{-1}\left({{r}\over{69 {\rm au}}}\right)^{3/2}  \, ,
\end{eqnarray}
where we have taken $\sigma_h = h$, and have defined metallicity $Z = \Sigma_{\rm dust}/\Sigma_{\rm gas}$. The number density $n \sim f {{\Sigma_d/h} \over{\rho_{\rm bulk} (4\pi/3) a^3}}$, where $f$ is the mass fraction of the small grains (we find $f \sim 1$), and the collisional cross-section $\sigma = \pi a^2$. 
We have also equated the relative collisional velocity $\Delta V$ to the vertical settling speed at a scale height, $v_z = c_s\, St$. 
This choice makes the collision timescale independent of grain size. The presence of any other velocity dispersion will further shorten the timescale. 

In fact, past theoretical studies have invariably concluded that the dust mass should be quickly concentrated into large grains \citep[e.g.,][]{Birnstiel2012}. We confirm such claims using the 
DustPy code \citep{Stammler2022} and present the results in Appendix \ref{sec:dustpy}.
In a LkCa-15-like ring, grains starting from $1\mu m$ quickly conglomerate to cm-sized after a mere 0.2 Myrs. 
As a result, the ring should appear both narrow and thin, with, e.g., scale heights $\sigma_h/r \sim 0.02-0.01$ going from Band 7 to 3 (Fig. \ref{fig:plot_dustpy2}).
These grains may resemble the `big grains' in LkCa 15, but fail to explain the broad and puffy ring of `small grains'.

The LkCa 15 ring may not be unique. 
At least two other systems, HD 163296 and V1094Sco, are reported to host broad and thick  dust rings \citep{Doi2021,Villenave2025}.
They may, like LkCa 15, also possess a large amount of small grains. 

Why are the small grains so massive? One possibility is that they are prevented to grow by the presence of a bouncing barrier \citep{Dominik1997,Dominik2024,Qian2025}.

Another puzzle relates to the absence of the so-called `hot-wall'. The 69au ring is inclined. So, we can see the star-facing wall on the far side of the ring (along one minor axis).
One expects this wall to be  markedly hotter (and therefore brighter) than the near-side which faces away from the star. Such hot-wall effect have been reported in disks like CIDA 9, RY Tau \citep{Ribas2024}, as well as in the inner region of HL Tau \citep{Guerra-Alvarado2024}, and has been used to measure disk thickness. But surprisingly, it is absent in LkCa 15 -- both its minor axis are largely similar in brightness. This is surprising in Band 7 which has a large optical depth, and even more so in Band 9 \citep[see image in][]{Leemker2022} where the optical depth is likely even higher. Such an absence of the `hot-wall' has also been reported for a number of well resolved rings, e.g., GM Aur \citep{Huang2020} and J16083070 \citep{Villenave2025}. 
It is unclear what causes this.

Now we turn to caveats. {First, our measurements of the ring images may carry some systematic errors that are introduced by, e.g., neighbouring rings, elliptical beam shapes, low signal-to-noise ratio (especially in Band 3). Higher fidelity images will help confirm our results.}
Our model also contains a number of  simplifications. One is the assumption of a Gaussian ring. This can be refined by fitting the radial profile along the major axis \citep[as is done in][]{Villenave2025}. This assumption is also suspicious in the context of the two-size model, since the composite of two Gaussians may not be well fit by a Gaussian. The two-size model itself can also be improved upon when further constraints arrive.

Throughout this work, we ignore the effects of grain scattering. Large grains, on the other hand, may have high albedos in ALMA bands (Fig. \ref{fig:diana}). In Appendix \ref{sec:scatter}, we solicit the help of the RADMC3d code \citep{radmc} to help understand how much our results are altered by grain scattering. Fortuitously, for the parameters of the 69au disk (width and height), a ring with high albedo appears the same as one with zero albedo. The optical depths that we infer for the ring can be straightforwardly interpreted as the total optical depths values (scattering + absorption). This conclusion may not hold for other rings. 

Our geometrical method may also be plagued by the presence of intrinsic asymmetries, e.g., those caused by shadows. The latter may explain the imperfect brightness fit in Fig. \ref{fig:phiobs}.

\section{Conclusions}
\label{sec:conclusion}

\citetalias{Doi2021} first pointed out that the brightness asymmetry of an inclined dust ring can be used to measure its vertical height. In this work, we extend their idea beyond the original optically thin limit, and beyond the sole use of brightness asymmetry. We demonstrate that, when one also measures the apparent radial width along the azimuth, three independent quantities can be constrained simultaneously: ring optical depth, intrinsic width and intrinsic height. However, to do so requires knowing the exact beam shape. 
Our method is, as is the original \citetalias{Doi2021} work, purely geometrical. We are thus free from assumptions on disk temperature and  opacity laws. 

Applying our method to the 69au ring of LkCa 15, we find that it appears broad and thick. In ALMA Band 7, its fractional radial width and vertical thickness are both of order $10\%$. Conventional wisdom holds that ALMA emission is dominated by large dust grains. But if so, such a fluffy ring would demand un-physically strong stirring.

Fortunately, the extensive data on LkCa 15 allows us to also study the ring in both Bands 6 and 3. We find that the ring width and height smoothly decrease with increasing wavelength. A single dust population would not appear to have different intrinsic geometry in different bands. So our results demand not one, but two or more dust populations.  These populations, with their different opacity law and different spatial distributions, trade dominance in different ALMA bands.

To explore this idea, we experiment with a simple two-size model for the 69au ring. Our model suggests that the `small' grain population (likely anything smaller than $\sim 20\mu$m) are spatially broad and thick; while the `big' population ($\sim 200\mu m$ and above) are distributed a few times narrower and thinner.  
This latter fact relieves the need to invoke strong stirring. We also find that the small population dominates the flux in Band 7 and dominates the total solid budget. These conclusions depart from conventional wisdom (e.g., DustPy simulations) and are surprising.

Is LkCa 15 special? Compared to other bright disks in Taurus, the spectral energy distribution of LkCa 15 appears unremarkable \citep[see, e.g.,][]{Painter2025}. 
Moreover, among dusty rings studied for their heights, some are reported to be as vertically extended \citep[typically in Band 6, see e.g.,][]{Villenave2025,Antilen2026}. 
So we speculate that LkCa 15 is not special, and that  a substantial fraction of the disks also have large  masses of small grains. This can be confirmed by applying our geometrical method on future multi-band, high-resolution data.

\bigskip
\bigskip

(Acknowledgement to be inserted)

\bigskip

\bibliography{ref}

@ARTICLE{Birnstiel2012,
       author = {{Birnstiel}, T. and {Klahr}, H. and {Ercolano}, B.},
        title = "{A simple model for the evolution of the dust population in protoplanetary disks}",
      journal = {\aap},
         year = 2012,
        month = mar,
       volume = {539},
          eid = {A148},
        pages = {A148},
          doi = {10.1051/0004-6361/201118136},
archivePrefix = {arXiv},
       eprint = {1201.5781},
 primaryClass = {astro-ph.EP},
       adsurl = {https://ui.adsabs.harvard.edu/abs/2012A&A...539A.148B}
}

@ARTICLE{Xu2022,
       author = {{Xu}, Ziyan and {Bai}, Xue-Ning},
        title = "{Dust Settling and Clumping in MRI-turbulent Outer Protoplanetary Disks}",
      journal = {\apj},
         year = 2022,
        month = jan,
       volume = {924},
       number = {1},
          eid = {3},
        pages = {3},
          doi = {10.3847/1538-4357/ac31a7},
archivePrefix = {arXiv},
       eprint = {2108.10486},
 primaryClass = {astro-ph.EP},
       adsurl = {https://ui.adsabs.harvard.edu/abs/2022ApJ...924....3X}
}

@ARTICLE{Dominik1997,
       author = {{Dominik}, C. and {Tielens}, A.~G.~G.~M.},
        title = "{The Physics of Dust Coagulation and the Structure of Dust Aggregates in Space}",
      journal = {\apj},
         year = 1997,
        month = may,
       volume = {480},
       number = {2},
        pages = {647-673},
          doi = {10.1086/303996},
       adsurl = {https://ui.adsabs.harvard.edu/abs/1997ApJ...480..647D}
}

@ARTICLE{Dominik2024,
       author = {{Dominik}, C. and {Dullemond}, C.~P.},
        title = "{The bouncing barrier revisited: Impact on key planet formation processes and observational signatures}",
      journal = {\aap},
         year = 2024,
        month = feb,
       volume = {682},
          eid = {A144},
        pages = {A144},
          doi = {10.1051/0004-6361/202347716},
archivePrefix = {arXiv},
       eprint = {2312.06000},
 primaryClass = {astro-ph.EP},
       adsurl = {https://ui.adsabs.harvard.edu/abs/2024A&A...682A.144D}
}

@ARTICLE{Leemker2022,
       author = {{Leemker}, M. and {Booth}, A.~S. and {van Dishoeck}, E.~F. and {P{\'e}rez-S{\'a}nchez}, A.~F. and {Szul{\'a}gyi}, J. and {Bosman}, A.~D. and {Bruderer}, S. and {Facchini}, S. and {Hogerheijde}, M.~R. and {Paneque-Carre{\~n}o}, T. and {Sturm}, J.~A.},
        title = "{Gas temperature structure across transition disk cavities}",
      journal = {\aap},
         year = 2022,
        month = jul,
       volume = {663},
          eid = {A23},
        pages = {A23},
          doi = {10.1051/0004-6361/202243229},
archivePrefix = {arXiv},
       eprint = {2204.03666},
 primaryClass = {astro-ph.EP},
       adsurl = {https://ui.adsabs.harvard.edu/abs/2022A&A...663A..23L}
}

@ARTICLE{Zhang2018,
       author = {{Zhang}, Shangjia and {Zhu}, Zhaohuan and {Huang}, Jane and {Guzm{\'a}n}, Viviana V. and {Andrews}, Sean M. and {Birnstiel}, Tilman and {Dullemond}, Cornelis P. and {Carpenter}, John M. and {Isella}, Andrea and {P{\'e}rez}, Laura M. and {Benisty}, Myriam and {Wilner}, David J. and {Baruteau}, Cl{\'e}ment and {Bai}, Xue-Ning and {Ricci}, Luca},
        title = "{The Disk Substructures at High Angular Resolution Project (DSHARP). VII. The Planet-Disk Interactions Interpretation}",
      journal = {\apjl},
         year = 2018,
        month = dec,
       volume = {869},
       number = {2},
          eid = {L47},
        pages = {L47},
          doi = {10.3847/2041-8213/aaf744},
archivePrefix = {arXiv},
       eprint = {1812.04045},
 primaryClass = {astro-ph.EP},
       adsurl = {https://ui.adsabs.harvard.edu/abs/2018ApJ...869L..47Z}
}

@ARTICLE{Long2018,
       author = {{Long}, Feng and {Pinilla}, Paola and {Herczeg}, Gregory J. and {Harsono}, Daniel and {Dipierro}, Giovanni and {Pascucci}, Ilaria and {Hendler}, Nathan and {Tazzari}, Marco and {Ragusa}, Enrico and {Salyk}, Colette and {Edwards}, Suzan and {Lodato}, Giuseppe and {van de Plas}, Gerrit and {Johnstone}, Doug and {Liu}, Yao and {Boehler}, Yann and {Cabrit}, Sylvie and {Manara}, Carlo F. and {Menard}, Francois and {Mulders}, Gijs D. and {Nisini}, Brunella and {Fischer}, William J. and {Rigliaco}, Elisabetta and {Banzatti}, Andrea and {Avenhaus}, Henning and {Gully-Santiago}, Michael},
        title = "{Gaps and Rings in an ALMA Survey of Disks in the Taurus Star-forming Region}",
      journal = {\apj},
         year = 2018,
        month = dec,
       volume = {869},
       number = {1},
          eid = {17},
        pages = {17},
          doi = {10.3847/1538-4357/aae8e1},
archivePrefix = {arXiv},
       eprint = {1810.06044},
 primaryClass = {astro-ph.SR},
       adsurl = {https://ui.adsabs.harvard.edu/abs/2018ApJ...869...17L}
}

@ARTICLE{AndrewsDSHARP,
       author = {{Andrews}, Sean M. and {Huang}, Jane and {P{\'e}rez}, Laura M. and {Isella}, Andrea and {Dullemond}, Cornelis P. and {Kurtovic}, Nicol{\'a}s T. and {Guzm{\'a}n}, Viviana V. and {Carpenter}, John M. and {Wilner}, David J. and {Zhang}, Shangjia and {Zhu}, Zhaohuan and {Birnstiel}, Tilman and {Bai}, Xue-Ning and {Benisty}, Myriam and {Hughes}, A. Meredith and {{\"O}berg}, Karin I. and {Ricci}, Luca},
        title = "{The Disk Substructures at High Angular Resolution Project (DSHARP). I. Motivation, Sample, Calibration, and Overview}",
      journal = {\apjl},
         year = 2018,
        month = dec,
       volume = {869},
       number = {2},
          eid = {L41},
        pages = {L41},
          doi = {10.3847/2041-8213/aaf741},
archivePrefix = {arXiv},
       eprint = {1812.04040},
 primaryClass = {astro-ph.SR},
       adsurl = {https://ui.adsabs.harvard.edu/abs/2018ApJ...869L..41A}
}

@ARTICLE{Painter2025,
       author = {{Painter}, Caleb and {Andrews}, Sean M. and {Chandler}, Claire J. and {Ueda}, Takahiro and {Wilner}, David J. and {Long}, Feng and {Macias}, Enrique and {Carrasco-Gonzalez}, Carlos and {Chung}, Chia-Ying and {Liu}, Hauyu Baobab and {Birnstiel}, Tilman and {Hughes}, A. Meredith},
        title = "{Detailed Microwave Continuum Spectra from Bright Protoplanetary Disks in Taurus}",
      journal = {The Open Journal of Astrophysics},
         year = 2025,
        month = sep,
       volume = {8},
          eid = {134},
        pages = {134},
          doi = {10.33232/001c.144268},
archivePrefix = {arXiv},
       eprint = {2507.21268},
 primaryClass = {astro-ph.SR},
       adsurl = {https://ui.adsabs.harvard.edu/abs/2025OJAp....8E.134P}
}

@ARTICLE{Stammler2022,
       author = {{Stammler}, Sebastian M. and {Birnstiel}, Tilman},
        title = "{DustPy: A Python Package for Dust Evolution in Protoplanetary Disks}",
      journal = {\apj},
         year = 2022,
        month = aug,
       volume = {935},
       number = {1},
          eid = {35},
        pages = {35},
          doi = {10.3847/1538-4357/ac7d58},
archivePrefix = {arXiv},
       eprint = {2207.00322},
 primaryClass = {astro-ph.EP},
       adsurl = {https://ui.adsabs.harvard.edu/abs/2022ApJ...935...35S}
}

@ARTICLE{Jin2019,
       author = {{Jin}, Sheng and {Isella}, Andrea and {Huang}, Pinghui and {Li}, Shengtai and {Li}, Hui and {Ji}, Jianghui},
        title = "{New Constraints on the Dust and Gas Distribution in the LkCa 15 Disk from ALMA}",
      journal = {\apj},
         year = 2019,
        month = aug,
       volume = {881},
       number = {2},
          eid = {108},
        pages = {108},
          doi = {10.3847/1538-4357/ab2dfe},
archivePrefix = {arXiv},
       eprint = {1907.00571},
 primaryClass = {astro-ph.EP},
       adsurl = {https://ui.adsabs.harvard.edu/abs/2019ApJ...881..108J}
}

@software{Optool,
       author = {{Dominik}, Carsten and {Min}, Michiel and {Tazaki}, Ryo},
        title = "{OpTool: Command-line driven tool for creating complex dust opacities}",
 howpublished = {Astrophysics Source Code Library, record ascl:2104.010},
         year = 2021,
        month = apr,
          eid = {ascl:2104.010},
archivePrefix = {ascl},
       eprint = {2104.010},
       adsurl = {https://ui.adsabs.harvard.edu/abs/2021ascl.soft04010D}
}

@ARTICLE{Woitke2016,
       author = {{Woitke}, P. and {Min}, M. and {Pinte}, C. and {Thi}, W.-F. and {Kamp}, I. and {Rab}, C. and {Anthonioz}, F. and {Antonellini}, S. and {Baldovin-Saavedra}, C. and {Carmona}, A. and {Dominik}, C. and {Dionatos}, O. and {Greaves}, J. and {G{\"u}del}, M. and {Ilee}, J.~D. and {Liebhart}, A. and {M{\'e}nard}, F. and {Rigon}, L. and {Waters}, L.~B.~F.~M. and {Aresu}, G. and {Meijerink}, R. and {Spaans}, M.},
        title = "{Consistent dust and gas models for protoplanetary disks. I. Disk shape, dust settling, opacities, and PAHs}",
      journal = {\aap},
         year = 2016,
        month = feb,
       volume = {586},
          eid = {A103},
        pages = {A103},
          doi = {10.1051/0004-6361/201526538},
archivePrefix = {arXiv},
       eprint = {1511.03431},
 primaryClass = {astro-ph.EP},
       adsurl = {https://ui.adsabs.harvard.edu/abs/2016A&A...586A.103W}
}

@ARTICLE{Swastik2026,
       author = {{Swastik}, C. and {Wahhaj}, Z. and {Benisty}, M. and {Arora}, S. and {Ginski}, C. and {Ren}, B.~B. and {van Holstein}, R.~G. and {de Rosa}, R. and {Banyal}, R.~K. and {Tazaki}, R.},
        title = "{Imaging the LkCa 15 system in polarimetry and total intensity without self-subtraction artefacts}",
      journal = {\aap},
         year = 2026,
        month = feb,
       volume = {706},
          eid = {A312},
        pages = {A312},
          doi = {10.1051/0004-6361/202449743},
archivePrefix = {arXiv},
       eprint = {2512.18439},
 primaryClass = {astro-ph.EP},
       adsurl = {https://ui.adsabs.harvard.edu/abs/2026A&A...706A.312S}
}

@ARTICLE{Sturm2023,
       author = {{Sturm}, J.~A. and {Booth}, A.~S. and {McClure}, M.~K. and {Leemker}, M. and {van Dishoeck}, E.~F.},
        title = "{Disentangling the protoplanetary disk gas mass and carbon depletion through combined atomic and molecular tracers}",
      journal = {\aap},
         year = 2023,
        month = feb,
       volume = {670},
          eid = {A12},
        pages = {A12},
          doi = {10.1051/0004-6361/202244227},
archivePrefix = {arXiv},
       eprint = {2209.09286},
 primaryClass = {astro-ph.EP},
       adsurl = {https://ui.adsabs.harvard.edu/abs/2023A&A...670A..12S}
}

@ARTICLE{Villenave2025,
       author = {{Villenave}, Marion and {Rosotti}, Giovanni P. and {Lambrechts}, Michiel and {Ziampras}, Alexandros and {Pinte}, Christophe and {M{\'e}nard}, Fran{\c{c}}ois and {Stapelfeldt}, Karl R. and {Duch{\^e}ne}, Gaspard and {Baylock}, Emily and {Doi}, Kiyoaki},
        title = "{Turbulence in protoplanetary disks: A systematic analysis of dust settling in 33 disks}",
      journal = {\aap},
         year = 2025,
        month = may,
       volume = {697},
          eid = {A64},
        pages = {A64},
          doi = {10.1051/0004-6361/202553822},
archivePrefix = {arXiv},
       eprint = {2503.05872},
 primaryClass = {astro-ph.SR},
       adsurl = {https://ui.adsabs.harvard.edu/abs/2025A&A...697A..64V}
}

@ARTICLE{Jiang2025,
       author = {{Jiang}, Haochang and {Long}, Feng and {Mac{\'\i}as}, Enrique and {Benisty}, Myriam and {Doi}, Kiyoaki and {Dullemond}, Cornelis P. and {Loomis}, Ryan A. and {Pascucci}, Ilaria and {P{\'e}rez}, Sebasti{\'a}n and {Zhang}, Shangjia and {Zhu}, Zhaohuan},
        title = "{Puffed-up Inner Rings and Razor-thin Outer Rings in Structured Protoplanetary Disks}",
      journal = {\apj},
         year = 2025,
        month = nov,
       volume = {993},
       number = {2},
          eid = {166},
        pages = {166},
          doi = {10.3847/1538-4357/ae089e},
archivePrefix = {arXiv},
       eprint = {2509.13122},
 primaryClass = {astro-ph.EP},
       adsurl = {https://ui.adsabs.harvard.edu/abs/2025ApJ...993..166J}
}

@ARTICLE{Doi2021,
       author = {{Doi}, Kiyoaki and {Kataoka}, Akimasa},
        title = "{Estimate on Dust Scale Height from the ALMA Dust Continuum Image of the HD 163296 Protoplanetary Disk}",
      journal = {\apj},
         year = 2021,
        month = may,
       volume = {912},
       number = {2},
          eid = {164},
        pages = {164},
          doi = {10.3847/1538-4357/abe5a6},
archivePrefix = {arXiv},
       eprint = {2102.06209},
 primaryClass = {astro-ph.EP},
       adsurl = {https://ui.adsabs.harvard.edu/abs/2021ApJ...912..164D}
}

@ARTICLE{Dullemond2004,
       author = {{Dullemond}, C.~P. and {Dominik}, C.},
        title = "{The effect of dust settling on the appearance  of protoplanetary disks}",
      journal = {\aap},
         year = 2004,
        month = jul,
       volume = {421},
        pages = {1075-1086},
          doi = {10.1051/0004-6361:20040284},
archivePrefix = {arXiv},
       eprint = {astro-ph/0405226},
 primaryClass = {astro-ph},
       adsurl = {https://ui.adsabs.harvard.edu/abs/2004A&A...421.1075D}
}

@ARTICLE{Tazzari2021,
       author = {{Tazzari}, M. and {Testi}, L. and {Natta}, A. and {Williams}, J.~P. and {Ansdell}, M. and {Carpenter}, J.~M. and {Facchini}, S. and {Guidi}, G. and {Hogherheijde}, M. and {Manara}, C.~F. and {Miotello}, A. and {van der Marel}, N.},
        title = "{The first ALMA survey of protoplanetary discs at 3 mm: demographics of grain growth in the Lupus region}",
      journal = {\mnras},
         year = 2021,
        month = oct,
       volume = {506},
       number = {4},
        pages = {5117-5128},
          doi = {10.1093/mnras/stab1912},
archivePrefix = {arXiv},
       eprint = {2010.02248},
 primaryClass = {astro-ph.EP},
       adsurl = {https://ui.adsabs.harvard.edu/abs/2021MNRAS.506.5117T}
}

@ARTICLE{Ribas2024,
       author = {{Ribas}, {\'A}. and {Clarke}, Cathie J. and {Zagaria}, Francesco},
        title = "{Inner walls or vortices? Crescent-shaped asymmetries in ALMA observations of protoplanetary discs}",
      journal = {\mnras},
         year = 2024,
        month = aug,
       volume = {532},
       number = {2},
        pages = {1752-1764},
          doi = {10.1093/mnras/stae1534},
archivePrefix = {arXiv},
       eprint = {2406.14626},
 primaryClass = {astro-ph.EP},
       adsurl = {https://ui.adsabs.harvard.edu/abs/2024MNRAS.532.1752R}
}

@ARTICLE{Pinte2016,
       author = {{Pinte}, C. and {Dent}, W.~R.~F. and {M{\'e}nard}, F. and {Hales}, A. and {Hill}, T. and {Cortes}, P. and {de Gregorio-Monsalvo}, I.},
        title = "{Dust and Gas in the Disk of HL Tauri: Surface Density, Dust Settling, and Dust-to-gas Ratio}",
      journal = {\apj},
         year = 2016,
        month = jan,
       volume = {816},
       number = {1},
          eid = {25},
        pages = {25},
          doi = {10.3847/0004-637X/816/1/25},
archivePrefix = {arXiv},
       eprint = {1508.00584},
 primaryClass = {astro-ph.SR},
       adsurl = {https://ui.adsabs.harvard.edu/abs/2016ApJ...816...25P}
}

@ARTICLE{Villenave2022,
       author = {{Villenave}, M. and {Stapelfeldt}, K.~R. and {Duch{\^e}ne}, G. and {M{\'e}nard}, F. and {Lambrechts}, M. and {Sierra}, A. and {Flores}, C. and {Dent}, W.~R.~F. and {Wolff}, S. and {Ribas}, {\'A}. and {Benisty}, M. and {Cuello}, N. and {Pinte}, C.},
        title = "{A Highly Settled Disk around Oph163131}",
      journal = {\apj},
         year = 2022,
        month = may,
       volume = {930},
       number = {1},
          eid = {11},
        pages = {11},
          doi = {10.3847/1538-4357/ac5fae},
archivePrefix = {arXiv},
       eprint = {2204.00640},
 primaryClass = {astro-ph.SR},
       adsurl = {https://ui.adsabs.harvard.edu/abs/2022ApJ...930...11V}
}

@ARTICLE{Bai2013,
       author = {{Bai}, Xue-Ning and {Stone}, James M.},
        title = "{Wind-driven Accretion in Protoplanetary Disks. I. Suppression of the Magnetorotational Instability and Launching of the Magnetocentrifugal Wind}",
      journal = {\apj},
         year = 2013,
        month = may,
       volume = {769},
       number = {1},
          eid = {76},
        pages = {76},
          doi = {10.1088/0004-637X/769/1/76},
archivePrefix = {arXiv},
       eprint = {1301.0318},
 primaryClass = {astro-ph.EP},
       adsurl = {https://ui.adsabs.harvard.edu/abs/2013ApJ...769...76B}
}

@ARTICLE{Guerra-Alvarado2024,
       author = {{Guerra-Alvarado}, Osmar M. and {Carrasco-Gonz{\'a}lez}, Carlos and {Mac{\'\i}as}, Enrique and {van der Marel}, Nienke and {Houge}, Adrien and {Maud}, Luke T. and {Pinilla}, Paola and {Villenave}, Marion and {Asaki}, Yoshiharu and {Humphreys}, Elizabeth},
        title = "{Into the thick of it: ALMA 0.45 mm observations of HL Tau at a resolution of 2 au}",
      journal = {\aap},
         year = 2024,
        month = jun,
       volume = {686},
          eid = {A298},
        pages = {A298},
          doi = {10.1051/0004-6361/202349046},
archivePrefix = {arXiv},
       eprint = {2404.04164},
 primaryClass = {astro-ph.EP},
       adsurl = {https://ui.adsabs.harvard.edu/abs/2024A&A...686A.298G}
}

@ARTICLE{Villenave2025review,
       author = {{Villenave}, Marion},
        title = "{An Introduction to Dust Evolution and Vertical Transport in Protoplanetary Disks}",
      journal = {\pasp},
         year = 2025,
        month = oct,
       volume = {137},
       number = {10},
          eid = {103001},
        pages = {103001},
          doi = {10.1088/1538-3873/ae05cb},
archivePrefix = {arXiv},
       eprint = {2509.10614},
 primaryClass = {astro-ph.SR},
       adsurl = {https://ui.adsabs.harvard.edu/abs/2025PASP..137j3001V}
}

@ARTICLE{Andrews2018,
       author = {{Andrews}, Sean M. and {Terrell}, Marie and {Tripathi}, Anjali and {Ansdell}, Megan and {Williams}, Jonathan P. and {Wilner}, David J.},
        title = "{Scaling Relations Associated with Millimeter Continuum Sizes in Protoplanetary Disks}",
      journal = {\apj},
         year = 2018,
        month = oct,
       volume = {865},
       number = {2},
          eid = {157},
        pages = {157},
          doi = {10.3847/1538-4357/aadd9f},
archivePrefix = {arXiv},
       eprint = {1808.10510},
 primaryClass = {astro-ph.EP},
       adsurl = {https://ui.adsabs.harvard.edu/abs/2018ApJ...865..157A}
}

@ARTICLE{Tripathi2017,
       author = {{Tripathi}, Anjali and {Andrews}, Sean M. and {Birnstiel}, Tilman and {Wilner}, David J.},
        title = "{A millimeter Continuum Size-Luminosity Relationship for Protoplanetary Disks}",
      journal = {\apj},
         year = 2017,
        month = aug,
       volume = {845},
       number = {1},
          eid = {44},
        pages = {44},
          doi = {10.3847/1538-4357/aa7c62},
archivePrefix = {arXiv},
       eprint = {1706.08977},
 primaryClass = {astro-ph.EP},
       adsurl = {https://ui.adsabs.harvard.edu/abs/2017ApJ...845...44T}
}

@ARTICLE{Qian2025,
       author = {{Qian}, Yansong and {Wu}, Yanqin},
        title = "{Bouncing Grains Keep Protoplanetary Disks Bright}",
      journal = {\apj},
         year = 2025,
        month = dec,
       volume = {995},
       number = {1},
          eid = {115},
        pages = {115},
          doi = {10.3847/1538-4357/ae17c5},
archivePrefix = {arXiv},
       eprint = {2507.06298},
 primaryClass = {astro-ph.EP},
       adsurl = {https://ui.adsabs.harvard.edu/abs/2025ApJ...995..115Q}
}

@ARTICLE{Otter2021,
       author = {{Otter}, Justin and {Ginsburg}, Adam and {Ballering}, Nicholas P. and {Bally}, John and {Eisner}, J.~A. and {Goddi}, Ciriaco and {Plambeck}, Richard and {Wright}, Melvyn},
        title = "{Small Protoplanetary Disks in the Orion Nebula Cluster and OMC1 with ALMA}",
      journal = {\apj},
         year = 2021,
        month = dec,
       volume = {923},
       number = {2},
          eid = {221},
        pages = {221},
          doi = {10.3847/1538-4357/ac29c2},
archivePrefix = {arXiv},
       eprint = {2109.14592},
 primaryClass = {astro-ph.GA},
       adsurl = {https://ui.adsabs.harvard.edu/abs/2021ApJ...923..221O}
}

@ARTICLE{Barenfeld2017,
       author = {{Barenfeld}, Scott A. and {Carpenter}, John M. and {Sargent}, Anneila I. and {Isella}, Andrea and {Ricci}, Luca},
        title = "{Measurement of Circumstellar Disk Sizes in the Upper Scorpius OB Association with ALMA}",
      journal = {\apj},
         year = 2017,
        month = dec,
       volume = {851},
       number = {2},
          eid = {85},
        pages = {85},
          doi = {10.3847/1538-4357/aa989d},
archivePrefix = {arXiv},
       eprint = {1711.04045},
 primaryClass = {astro-ph.SR},
       adsurl = {https://ui.adsabs.harvard.edu/abs/2017ApJ...851...85B}
}

@ARTICLE{Pinilla2014,
       author = {{Pinilla}, P. and {Benisty}, M. and {Birnstiel}, T. and {Ricci}, L. and {Isella}, A. and {Natta}, A. and {Dullemond}, C.~P. and {Quiroga-Nu{\~n}ez}, L.~H. and {Henning}, T. and {Testi}, L.},
        title = "{Millimetre spectral indices of transition disks and their relation to the cavity radius}",
      journal = {\aap},
         year = 2014,
        month = apr,
       volume = {564},
          eid = {A51},
        pages = {A51},
          doi = {10.1051/0004-6361/201323322},
archivePrefix = {arXiv},
       eprint = {1402.5778},
 primaryClass = {astro-ph.EP},
       adsurl = {https://ui.adsabs.harvard.edu/abs/2014A&A...564A..51P}
}

@ARTICLE{Dubrulle1995,
       author = {{Dubrulle}, B. and {Morfill}, G. and {Sterzik}, M.},
        title = "{The dust subdisk in the protoplanetary nebula.}",
      journal = {\icarus},
         year = 1995,
        month = apr,
       volume = {114},
       number = {2},
        pages = {237-246},
          doi = {10.1006/icar.1995.1058},
       adsurl = {https://ui.adsabs.harvard.edu/abs/1995Icar..114..237D}
}

@ARTICLE{Youdin2007,
       author = {{Youdin}, Andrew N. and {Lithwick}, Yoram},
        title = "{Particle stirring in turbulent gas disks: Including orbital oscillations}",
      journal = {\icarus},
         year = 2007,
        month = dec,
       volume = {192},
       number = {2},
        pages = {588-604},
          doi = {10.1016/j.icarus.2007.07.012},
archivePrefix = {arXiv},
       eprint = {0707.2975},
 primaryClass = {astro-ph},
       adsurl = {https://ui.adsabs.harvard.edu/abs/2007Icar..192..588Y}
}

@ARTICLE{Sierra2025,
       author = {{Sierra}, Anibal and {Pinilla}, Paola and {P{\'e}rez}, Laura M. and {Benisty}, Myriam and {Agurto-Gangas}, Carolina and {Carrasco-Gonz{\'a}lez}, Carlos and {Curone}, Pietro and {Long}, Feng},
        title = "{High angular resolution evidence of dust traps from deep ALMA Band 3 observations of LkCa15}",
      journal = {\mnras},
         year = 2025,
        month = apr,
       volume = {538},
       number = {4},
        pages = {2358-2374},
          doi = {10.1093/mnras/staf393},
archivePrefix = {arXiv},
       eprint = {2503.03336},
 primaryClass = {astro-ph.EP},
       adsurl = {https://ui.adsabs.harvard.edu/abs/2025MNRAS.538.2358S}
}

@ARTICLE{Pizzati2023,
       author = {{Pizzati}, Elia and {Rosotti}, Giovanni P. and {Tabone}, Beno{\^\i}t},
        title = "{Constraining turbulence in protoplanetary discs using the gap contrast: an application to the DSHARP sample}",
      journal = {\mnras},
         year = 2023,
        month = sep,
       volume = {524},
       number = {2},
        pages = {3184-3200},
          doi = {10.1093/mnras/stad2057},
archivePrefix = {arXiv},
       eprint = {2307.11150},
 primaryClass = {astro-ph.EP},
       adsurl = {https://ui.adsabs.harvard.edu/abs/2023MNRAS.524.3184P}
}

@ARTICLE{Villenave2020,
       author = {{Villenave}, M. and {M{\'e}nard}, F. and {Dent}, W.~R.~F. and {Duch{\^e}ne}, G. and {Stapelfeldt}, K.~R. and {Benisty}, M. and {Boehler}, Y. and {van der Plas}, G. and {Pinte}, C. and {Telkamp}, Z. and {Wolff}, S. and {Flores}, C. and {Lesur}, G. and {Louvet}, F. and {Riols}, A. and {Dougados}, C. and {Williams}, H. and {Padgett}, D.},
        title = "{Observations of edge-on protoplanetary disks with ALMA. I. Results from continuum data}",
      journal = {\aap},
         year = 2020,
        month = oct,
       volume = {642},
          eid = {A164},
        pages = {A164},
          doi = {10.1051/0004-6361/202038087},
archivePrefix = {arXiv},
       eprint = {2008.06518},
 primaryClass = {astro-ph.SR},
       adsurl = {https://ui.adsabs.harvard.edu/abs/2020A&A...642A.164V}
}

@ARTICLE{Huang2020,
       author = {{Huang}, Jane and {Andrews}, Sean M. and {Dullemond}, Cornelis P. and {{\"O}berg}, Karin I. and {Qi}, Chunhua and {Zhu}, Zhaohuan and {Birnstiel}, Tilman and {Carpenter}, John M. and {Isella}, Andrea and {Mac{\'\i}as}, Enrique and {McClure}, Melissa K. and {P{\'e}rez}, Laura M. and {Teague}, Richard and {Wilner}, David J. and {Zhang}, Shangjia},
        title = "{A Multifrequency ALMA Characterization of Substructures in the GM Aur Protoplanetary Disk}",
      journal = {\apj},
         year = 2020,
        month = mar,
       volume = {891},
       number = {1},
          eid = {48},
        pages = {48},
          doi = {10.3847/1538-4357/ab711e},
archivePrefix = {arXiv},
       eprint = {2001.11040},
 primaryClass = {astro-ph.EP},
       adsurl = {https://ui.adsabs.harvard.edu/abs/2020ApJ...891...48H}
}

@ARTICLE{Dullemond2018,
       author = {{Dullemond}, Cornelis P. and {Birnstiel}, Tilman and {Huang}, Jane and {Kurtovic}, Nicol{\'a}s T. and {Andrews}, Sean M. and {Guzm{\'a}n}, Viviana V. and {P{\'e}rez}, Laura M. and {Isella}, Andrea and {Zhu}, Zhaohuan and {Benisty}, Myriam and {Wilner}, David J. and {Bai}, Xue-Ning and {Carpenter}, John M. and {Zhang}, Shangjia and {Ricci}, Luca},
        title = "{The Disk Substructures at High Angular Resolution Project (DSHARP). VI. Dust Trapping in Thin-ringed Protoplanetary Disks}",
      journal = {\apjl},
         year = 2018,
        month = dec,
       volume = {869},
       number = {2},
          eid = {L46},
        pages = {L46},
          doi = {10.3847/2041-8213/aaf742},
archivePrefix = {arXiv},
       eprint = {1812.04044},
 primaryClass = {astro-ph.EP},
       adsurl = {https://ui.adsabs.harvard.edu/abs/2018ApJ...869L..46D}
}

@ARTICLE{Martinien2026,
       author = {{Martinien}, L. and {Duch{\^e}ne}, G. and {Ribas}, {\'A}. and {Villenave}, M. and {M{\'e}nard}, F. and {Stapelfeldt}, K.~R. and {Pinte}, C.},
        title = "{Observations of highly inclined disks with ALMA: Results from $^{12}$CO gas and continuum observations}",
      journal = {\aap},
         year = 2026,
        month = jun,
       volume = {710},
          eid = {A140},
        pages = {A140},
          doi = {10.1051/0004-6361/202659107},
archivePrefix = {arXiv},
       eprint = {2604.11247},
 primaryClass = {astro-ph.SR},
       adsurl = {https://ui.adsabs.harvard.edu/abs/2026A&A...710A.140M}
}

@ARTICLE{Long2022,
       author = {{Long}, Feng and {Andrews}, Sean M. and {Zhang}, Shangjia and {Qi}, Chunhua and {Benisty}, Myriam and {Facchini}, Stefano and {Isella}, Andrea and {Wilner}, David J. and {Bae}, Jaehan and {Huang}, Jane and {Loomis}, Ryan A. and {{\"O}berg}, Karin I. and {Zhu}, Zhaohuan},
        title = "{ALMA Detection of Dust Trapping around Lagrangian Points in the LkCa 15 Disk}",
      journal = {\apjl},
         year = 2022,
        month = sep,
       volume = {937},
       number = {1},
          eid = {L1},
        pages = {L1},
          doi = {10.3847/2041-8213/ac8b10},
archivePrefix = {arXiv},
       eprint = {2209.05535},
 primaryClass = {astro-ph.EP},
       adsurl = {https://ui.adsabs.harvard.edu/abs/2022ApJ...937L...1L}
}

@ARTICLE{Facchini2020,
       author = {{Facchini}, S. and {Benisty}, M. and {Bae}, J. and {Loomis}, R. and {Perez}, L. and {Ansdell}, M. and {Mayama}, S. and {Pinilla}, P. and {Teague}, R. and {Isella}, A. and {Mann}, A.},
        title = "{Annular substructures in the transition disks around LkCa 15 and J1610}",
      journal = {\aap},
         year = 2020,
        month = jul,
       volume = {639},
          eid = {A121},
        pages = {A121},
          doi = {10.1051/0004-6361/202038027},
archivePrefix = {arXiv},
       eprint = {2005.02712},
 primaryClass = {astro-ph.EP},
       adsurl = {https://ui.adsabs.harvard.edu/abs/2020A&A...639A.121F}
}

@ARTICLE{Gaia2018,
       author = {{Gaia Collaboration} and {Brown}, A.~G.~A. and {Vallenari}, A. and {Prusti}, T. and {de Bruijne}, J.~H.~J. and {Babusiaux}, C. and {Bailer-Jones}, C.~A.~L. and {Biermann}, M. and {Evans}, D.~W. and {Eyer}, L. and {Jansen}, F. and {Jordi}, C. and {Klioner}, S.~A. and {Lammers}, U. and {Lindegren}, L. and {Luri}, X. and {Mignard}, F. and {Panem}, C. and {Pourbaix}, D. and {Randich}, S. and {Sartoretti}, P. and {Siddiqui}, H.~I. and {Soubiran}, C. and {van Leeuwen}, F. and {Walton}, N.~A. and {Arenou}, F. and {Bastian}, U. and {Cropper}, M. and {Drimmel}, R. and {Katz}, D. and {Lattanzi}, M.~G. and {Bakker}, J. and {Cacciari}, C. and {Casta{\~n}eda}, J. and {Chaoul}, L. and {Cheek}, N. and {De Angeli}, F. and {Fabricius}, C. and {Guerra}, R. and {Holl}, B. and {Masana}, E. and {Messineo}, R. and {Mowlavi}, N. and {Nienartowicz}, K. and {Panuzzo}, P. and {Portell}, J. and {Riello}, M. and {Seabroke}, G.~M. and {Tanga}, P. and {Th{\'e}venin}, F. and {Gracia-Abril}, G. and {Comoretto}, G. and {Garcia-Reinaldos}, M. and {Teyssier}, D. and {Altmann}, M. and {Andrae}, R. and {Audard}, M. and {Bellas-Velidis}, I. and {Benson}, K. and {Berthier}, J. and {Blomme}, R. and {Burgess}, P. and {Busso}, G. and {Carry}, B. and {Cellino}, A. and {Clementini}, G. and {Clotet}, M. and {Creevey}, O. and {Davidson}, M. and {De Ridder}, J. and {Delchambre}, L. and {Dell'Oro}, A. and {Ducourant}, C. and {Fern{\'a}ndez-Hern{\'a}ndez}, J. and {Fouesneau}, M. and {Fr{\'e}mat}, Y. and {Galluccio}, L. and {Garc{\'\i}a-Torres}, M. and {Gonz{\'a}lez-N{\'u}{\~n}ez}, J. and {Gonz{\'a}lez-Vidal}, J.~J. and {Gosset}, E. and {Guy}, L.~P. and {Halbwachs}, J.-L. and {Hambly}, N.~C. and {Harrison}, D.~L. and {Hern{\'a}ndez}, J. and {Hestroffer}, D. and {Hodgkin}, S.~T. and {Hutton}, A. and {Jasniewicz}, G. and {Jean-Antoine-Piccolo}, A. and {Jordan}, S. and {Korn}, A.~J. and {Krone-Martins}, A. and {Lanzafame}, A.~C. and {Lebzelter}, T. and {L{\"o}ffler}, W. and {Manteiga}, M. and {Marrese}, P.~M. and {Mart{\'\i}n-Fleitas}, J.~M. and {Moitinho}, A. and {Mora}, A. and {Muinonen}, K. and {Osinde}, J. and {Pancino}, E. and {Pauwels}, T. and {Petit}, J.-M. and {Recio-Blanco}, A. and {Richards}, P.~J. and {Rimoldini}, L. and {Robin}, A.~C. and {Sarro}, L.~M. and {Siopis}, C. and {Smith}, M. and {Sozzetti}, A. and {S{\"u}veges}, M. and {Torra}, J. and {van Reeven}, W. and {Abbas}, U. and {Abreu Aramburu}, A. and {Accart}, S. and {Aerts}, C. and {Altavilla}, G. and {{\'A}lvarez}, M.~A. and {Alvarez}, R. and {Alves}, J. and {Anderson}, R.~I. and {Andrei}, A.~H. and {Anglada Varela}, E. and {Antiche}, E. and {Antoja}, T. and {Arcay}, B. and {Astraatmadja}, T.~L. and {Bach}, N. and {Baker}, S.~G. and {Balaguer-N{\'u}{\~n}ez}, L. and {Balm}, P. and {Barache}, C. and {Barata}, C. and {Barbato}, D. and {Barblan}, F. and {Barklem}, P.~S. and {Barrado}, D. and {Barros}, M. and {Barstow}, M.~A. and {Bartholom{\'e} Mu{\~n}oz}, S. and {Bassilana}, J.-L. and {Becciani}, U. and {Bellazzini}, M. and {Berihuete}, A. and {Bertone}, S. and {Bianchi}, L. and {Bienaym{\'e}}, O. and {Blanco-Cuaresma}, S. and {Boch}, T. and {Boeche}, C. and {Bombrun}, A. and {Borrachero}, R. and {Bossini}, D. and {Bouquillon}, S. and {Bourda}, G. and {Bragaglia}, A. and {Bramante}, L. and {Breddels}, M.~A. and {Bressan}, A. and {Brouillet}, N. and {Br{\"u}semeister}, T. and {Brugaletta}, E. and {Bucciarelli}, B. and {Burlacu}, A. and {Busonero}, D. and {Butkevich}, A.~G. and {Buzzi}, R. and {Caffau}, E. and {Cancelliere}, R. and {Cannizzaro}, G. and {Cantat-Gaudin}, T. and {Carballo}, R. and {Carlucci}, T. and {Carrasco}, J.~M. and {Casamiquela}, L. and {Castellani}, M. and {Castro-Ginard}, A. and {Charlot}, P. and {Chemin}, L. and {Chiavassa}, A. and {Cocozza}, G. and {Costigan}, G. and {Cowell}, S. and {Crifo}, F. and {Crosta}, M. and {Crowley}, C. and {Cuypers}, J. and {Dafonte}, C. and {Damerdji}, Y. and {Dapergolas}, A. and {David}, P. and {David}, M. and {de Laverny}, P. and {De Luise}, F.},
        title = "{Gaia Data Release 2. Summary of the contents and survey properties}",
      journal = {\aap},
         year = 2018,
        month = aug,
       volume = {616},
          eid = {A1},
        pages = {A1},
          doi = {10.1051/0004-6361/201833051},
archivePrefix = {arXiv},
       eprint = {1804.09365},
 primaryClass = {astro-ph.GA},
       adsurl = {https://ui.adsabs.harvard.edu/abs/2018A&A...616A...1G}
}

@ARTICLE{Manara2014,
       author = {{Manara}, C.~F. and {Testi}, L. and {Natta}, A. and {Rosotti}, G. and {Benisty}, M. and {Ercolano}, B. and {Ricci}, L.},
        title = "{Gas content of transitional disks: a VLT/X-Shooter study of accretion and winds}",
      journal = {\aap},
         year = 2014,
        month = aug,
       volume = {568},
          eid = {A18},
        pages = {A18},
          doi = {10.1051/0004-6361/201323318},
archivePrefix = {arXiv},
       eprint = {1406.1428},
 primaryClass = {astro-ph.SR},
       adsurl = {https://ui.adsabs.harvard.edu/abs/2014A&A...568A..18M}
}

@ARTICLE{Birnstiel2018,
       author = {{Birnstiel}, Tilman and {Dullemond}, Cornelis P. and {Zhu}, Zhaohuan and {Andrews}, Sean M. and {Bai}, Xue-Ning and {Wilner}, David J. and {Carpenter}, John M. and {Huang}, Jane and {Isella}, Andrea and {Benisty}, Myriam and {P{\'e}rez}, Laura M. and {Zhang}, Shangjia},
        title = "{The Disk Substructures at High Angular Resolution Project (DSHARP). V. Interpreting ALMA Maps of Protoplanetary Disks in Terms of a Dust Model}",
      journal = {\apjl},
         year = 2018,
        month = dec,
       volume = {869},
       number = {2},
          eid = {L45},
        pages = {L45},
          doi = {10.3847/2041-8213/aaf743},
archivePrefix = {arXiv},
       eprint = {1812.04043},
 primaryClass = {astro-ph.SR},
       adsurl = {https://ui.adsabs.harvard.edu/abs/2018ApJ...869L..45B}
}

@ARTICLE{Chiang2010,
       author = {{Chiang}, E. and {Youdin}, A.~N.},
        title = "{Forming Planetesimals in Solar and Extrasolar Nebulae}",
      journal = {Annual Review of Earth and Planetary Sciences},
         year = 2010,
        month = may,
       volume = {38},
        pages = {493-522},
          doi = {10.1146/annurev-earth-040809-152513},
archivePrefix = {arXiv},
       eprint = {0909.2652},
 primaryClass = {astro-ph.EP},
       adsurl = {https://ui.adsabs.harvard.edu/abs/2010AREPS..38..493C}
}

@ARTICLE{Antilen2026,
       author = {{Antilen}, Juanita and {Pinilla}, Paola and {Li}, Dafa and {Villenave}, Marion and {Sierra}, Anibal and {Liu}, Yao and {Benisty}, Myriam and {Ginski}, Christian},
        title = "{Diverse dust vertical height and settling strength conditions in protoplanetary discs}",
      journal = {\mnras},
         year = 2026,
        month = may,
          doi = {10.1093/mnras/stag882},
archivePrefix = {arXiv},
       eprint = {2605.06904},
 primaryClass = {astro-ph.EP},
       adsurl = {https://ui.adsabs.harvard.edu/abs/2026MNRAS.tmp..906A}
}

@ARTICLE{ALMA2015,
       author = {{ALMA Partnership} and {Brogan}, C.~L. and {P{\'e}rez}, L.~M. and {Hunter}, T.~R. and {Dent}, W.~R.~F. and {Hales}, A.~S. and {Hills}, R.~E. and {Corder}, S. and {Fomalont}, E.~B. and {Vlahakis}, C. and {Asaki}, Y. and {Barkats}, D. and {Hirota}, A. and {Hodge}, J.~A. and {Impellizzeri}, C.~M.~V. and {Kneissl}, R. and {Liuzzo}, E. and {Lucas}, R. and {Marcelino}, N. and {Matsushita}, S. and {Nakanishi}, K. and {Phillips}, N. and {Richards}, A.~M.~S. and {Toledo}, I. and {Aladro}, R. and {Broguiere}, D. and {Cortes}, J.~R. and {Cortes}, P.~C. and {Espada}, D. and {Galarza}, F. and {Garcia-Appadoo}, D. and {Guzman-Ramirez}, L. and {Humphreys}, E.~M. and {Jung}, T. and {Kameno}, S. and {Laing}, R.~A. and {Leon}, S. and {Marconi}, G. and {Mignano}, A. and {Nikolic}, B. and {Nyman}, L.-A. and {Radiszcz}, M. and {Remijan}, A. and {Rod{\'o}n}, J.~A. and {Sawada}, T. and {Takahashi}, S. and {Tilanus}, R.~P.~J. and {Vila Vilaro}, B. and {Watson}, L.~C. and {Wiklind}, T. and {Akiyama}, E. and {Chapillon}, E. and {de Gregorio-Monsalvo}, I. and {Di Francesco}, J. and {Gueth}, F. and {Kawamura}, A. and {Lee}, C.-F. and {Nguyen Luong}, Q. and {Mangum}, J. and {Pietu}, V. and {Sanhueza}, P. and {Saigo}, K. and {Takakuwa}, S. and {Ubach}, C. and {van Kempen}, T. and {Wootten}, A. and {Castro-Carrizo}, A. and {Francke}, H. and {Gallardo}, J. and {Garcia}, J. and {Gonzalez}, S. and {Hill}, T. and {Kaminski}, T. and {Kurono}, Y. and {Liu}, H.-Y. and {Lopez}, C. and {Morales}, F. and {Plarre}, K. and {Schieven}, G. and {Testi}, L. and {Videla}, L. and {Villard}, E. and {Andreani}, P. and {Hibbard}, J.~E. and {Tatematsu}, K.},
        title = "{The 2014 ALMA Long Baseline Campaign: First Results from High Angular Resolution Observations toward the HL Tau Region}",
      journal = {\apjl},
         year = 2015,
        month = jul,
       volume = {808},
       number = {1},
          eid = {L3},
        pages = {L3},
          doi = {10.1088/2041-8205/808/1/L3},
archivePrefix = {arXiv},
       eprint = {1503.02649},
 primaryClass = {astro-ph.SR},
       adsurl = {https://ui.adsabs.harvard.edu/abs/2015ApJ...808L...3A}
}

@ARTICLE{Marel2013,
       author = {{van der Marel}, Nienke and {van Dishoeck}, Ewine F. and {Bruderer}, Simon and {Birnstiel}, Til and {Pinilla}, Paola and {Dullemond}, Cornelis P. and {van Kempen}, Tim A. and {Schmalzl}, Markus and {Brown}, Joanna M. and {Herczeg}, Gregory J. and {Mathews}, Geoffrey S. and {Geers}, Vincent},
        title = "{A Major Asymmetric Dust Trap in a Transition Disk}",
      journal = {Science},
         year = 2013,
        month = jun,
       volume = {340},
       number = {6137},
        pages = {1199-1202},
          doi = {10.1126/science.1236770},
archivePrefix = {arXiv},
       eprint = {1306.1768},
 primaryClass = {astro-ph.EP},
       adsurl = {https://ui.adsabs.harvard.edu/abs/2013Sci...340.1199V}
}

@software{radmc,
       author = {{Dullemond}, C.~P. and {Juhasz}, A. and {Pohl}, A. and {Sereshti}, F. and {Shetty}, R. and {Peters}, T. and {Commercon}, B. and {Flock}, M.},
        title = "{RADMC-3D: A multi-purpose radiative transfer tool}",
 howpublished = {Astrophysics Source Code Library, record ascl:1202.015},
         year = 2012,
        month = feb,
          eid = {ascl:1202.015},
archivePrefix = {ascl},
       eprint = {1202.015},
       adsurl = {https://ui.adsabs.harvard.edu/abs/2012ascl.soft02015D}
}

@ARTICLE{Doi2023,
       author = {{Doi}, Kiyoaki and {Kataoka}, Akimasa},
        title = "{Constraints on the Dust Size Distributions in the HD 163296 Disk from the Difference of the Apparent Dust Ring Widths between Two ALMA Bands}",
      journal = {\apj},
         year = 2023,
        month = nov,
       volume = {957},
       number = {1},
          eid = {11},
        pages = {11},
          doi = {10.3847/1538-4357/acf5df},
archivePrefix = {arXiv},
       eprint = {2308.16574},
 primaryClass = {astro-ph.EP},
       adsurl = {https://ui.adsabs.harvard.edu/abs/2023ApJ...957...11D}
}

@ARTICLE{emcee,
       author = {{Foreman-Mackey}, Daniel and {Hogg}, David W. and {Lang}, Dustin and {Goodman}, Jonathan},
        title = "{emcee: The MCMC Hammer}",
      journal = {\pasp},
         year = 2013,
        month = mar,
       volume = {125},
       number = {925},
        pages = {306},
          doi = {10.1086/670067},
archivePrefix = {arXiv},
       eprint = {1202.3665},
 primaryClass = {astro-ph.IM},
       adsurl = {https://ui.adsabs.harvard.edu/abs/2013PASP..125..306F}
}

\appendix

\section{MCMC results on the 69au Ring}

\begin{figure*}
    \centering
    \includegraphics[width=0.72\linewidth]{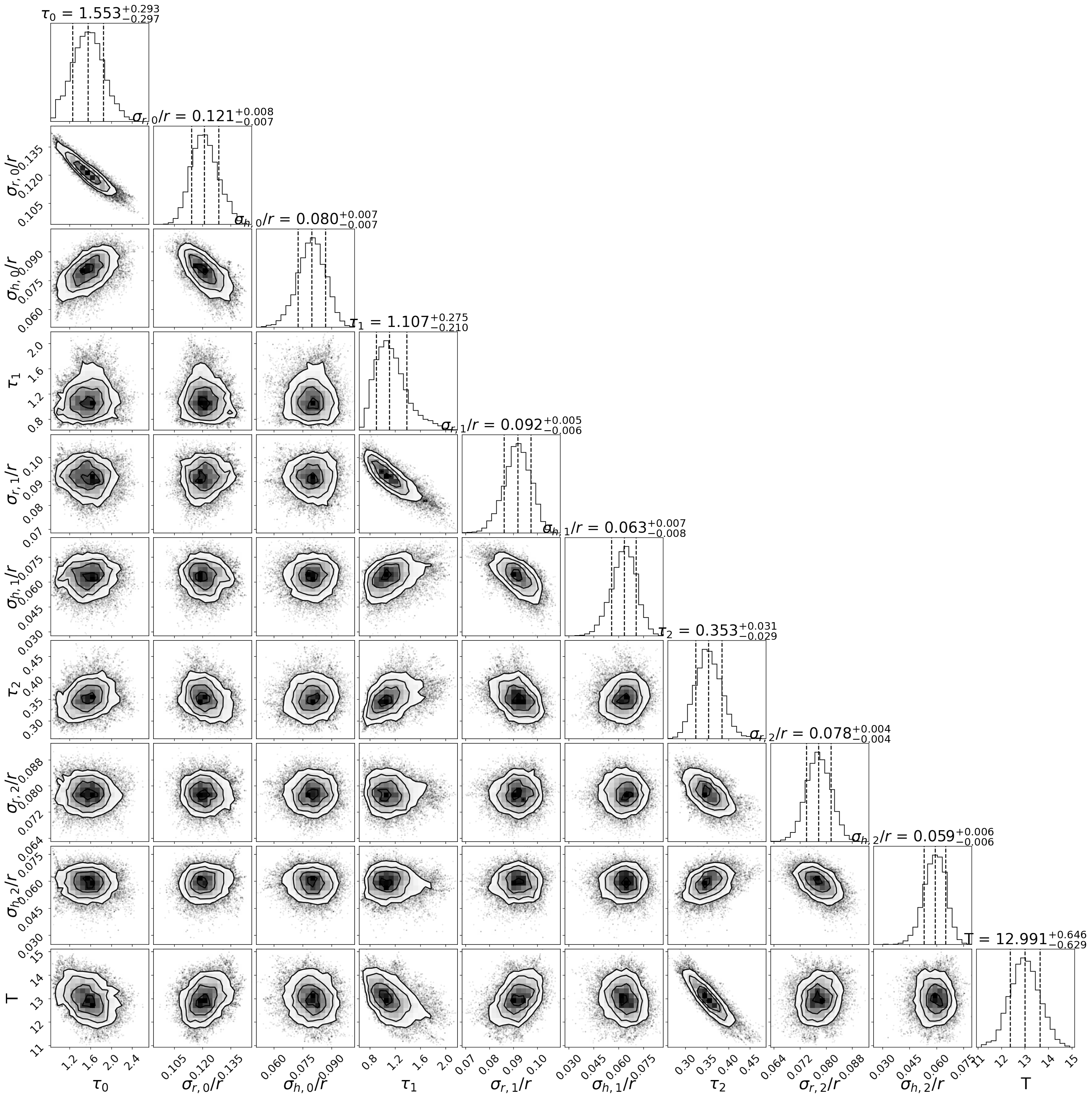}
    \caption{Multiwavelength (Bands 7/6/3) joint MCMC fitting results for the 69au ring.}
    \label{fig:corner}
\end{figure*}

Fig. \ref{fig:corner} displays the MCMC corner plot from the joint multi-wavelength (bands 7/6/3) fit for the 69au ring in LkCa 15. The best-fit from this exercise is listed in Table \ref{tab:fitting}. It is largely similar to the solutions from single-band fitting.

\section{The outer Ring}
\label{sec: outerring}
For completeness, we also carry out the same fitting procedure for the outer ring of LkCa 15 at 101AU. We only use the Band 6/7 data from \citet{Sierra2025} as the Band 3 data is too noisy. We use the same procedure as for the 69au ring, but limiting ourselves to regions outside the ring center. This is necessary to avoid contamination from the 69 au ring. The beam FWHM is $\sigma_B = 60$ mas ($\sigma_B/r = 0.041)$. The results are listed in Table \ref{tab:fitting-outer}.
As is the case for the 69au ring, ring optical depth also drops with increasing wavelength. The fractional radial widths are also similar, but the vertical heights appear smaller. In Band 6, we find $\sigma_h/r\sim 0.014$, consistent with  the upper limit of $\sigma_h/r \leq 0.034$ by \citet{Villenave2025}, but larger than the value of $\sigma_h/r \leq 0.006$ in \citet{Jiang2025}. These results are surprising as one naively expects a comparable width and height for a dust ring. One possibility is that it contains two unresolved thin rings. The other possibility relates to the residual noise pollution from the 69au ring.

\begin{table}[]
    \centering
    \begin{tabular}{c|cccc} 
    \hline
    \hline
  Band &\multicolumn{4}{c}{Best-fit} 
    \\ 
       & $T_{\rm ring}$ & $\tau_\perp$ & $\sigma_r/r$ & $\sigma_h/r$  
       \\
       \hline
7 
& $10.3^{+ 0.6 }_{- 0.5 }\K$ &$1.2^{+0.3}_{-0.2}$ &$0.136^{+ 0.005 }_{- 0.006 }$ &$0.034^{+ 0.007 }_{- 0.009 }$ 
\\
 & & & & \\
6 
&(same) 
&$0.59^{+ 0.09 }_{- 0.08 }$ &$0.131^{+ 0.002 }_{- 0.002 }$ &$0.014^{+ 0.004 }_{- 0.003 }$ 
\\
\hline   
\hline
    \end{tabular}
    \caption{ Joint fit for the outer (101au) ring of LkCa 15. The beam width $\sigma_B/r = 0.041$.}
    \label{tab:fitting-outer}
\end{table}

\section{Effects of Grain Scattering}
\label{sec:scatter}

\begin{figure}
    \centering
    \includegraphics[width=0.7\linewidth]{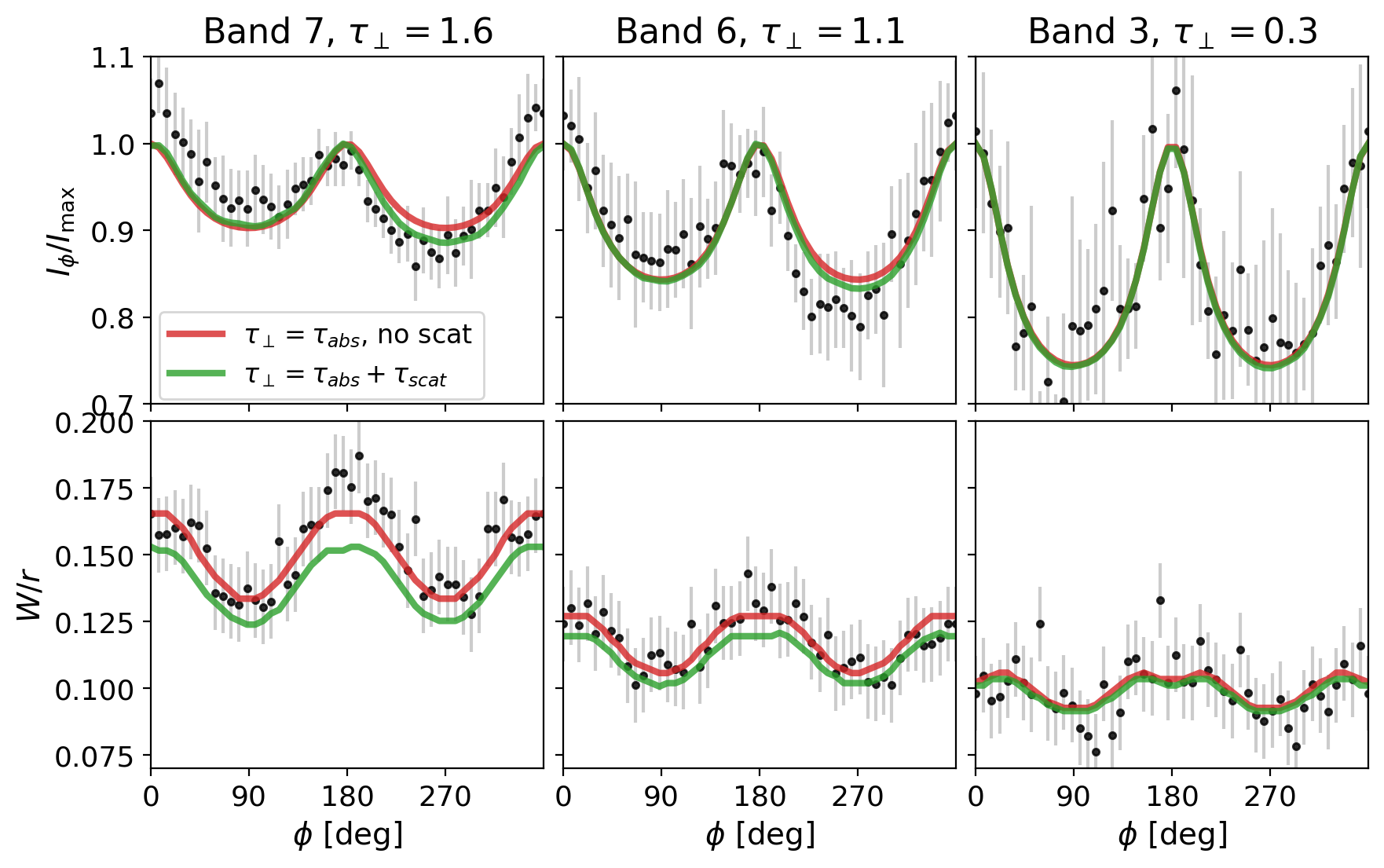}
    \caption{Effects of grain  scattering on the azimuthal profile of a dust ring. The data and the red curves (no scattering) are as in Fig. \ref{fig:phiobs},   
    while the green curves show the RADMC3d results (using the same set of $\sigma_h, \sigma_r$ and observed with the same beam) when the grain albedo is set to be $0.8$ and the peak optical depth refers to the total optical depth (absorption and scattering). We find that including scattering does not appreciably change the ring appearance, at least not for the geometrical parameters of the 69au ring.
    }
    \label{fig:scattering}
\end{figure}

Our analysis in the main paper ignores the effects of grain scattering. Here, we investigate if scattering can affect the azimuthal variations in disk brightness and with.

We model grain scattering using the Monte-Carlo radiative transfer code \texttt{RADMC-3D} \citep{radmc}. We set a constant albedo of $0.8$, or equivalently $\kappa_\mathrm{scat}=4\kappa_\mathrm{abs}$. This falls in the expected range for large grains (see  Fig. \ref{fig:diana}). Scattering is assumed to be isotropic. We further adopt the Multi-wavelength Joint Fit (Table \ref{tab:fitting}) for the dust spatial distribution and the peak optical depths. 

When the albedo is zero, the resultant azimuthal variations in each band are plotted in Fig. \ref{fig:phiobs} and reproduced here in Fig. \ref{fig:scattering}. When scattering is considered, we instead let the tabulated peak optical depth be the sum of absorption and scattering, $\tau_\perp = \tau_{\rm abs} + \tau_{\rm scat} $. This requires adjusting the dust density downwards by five times from the previous case. We then convolve the RADMC3d images with the observing beam ($\sigma_b/r = 0.059$), and extract the apparent brightness and width as in main text. These results are also plotted in Fig. \ref{fig:scattering}.

The scattering case, interestingly, looks almost identical to the no-scattering case, with a notice-able difference only in the most optically thick band (Band 7). So if we are to measure disk geometry using the no-scattering model (as is done in the main text), our results lie close to the truth, even with an albedo as high as $0.8$.
To be more precise, since the scattering model tends to make the ring look thinner (by about $15\%$ in Band 7), if we were to correct for the effect of scattering, we would have found that the intrinsic width in Band 7 is even higher than our current value. This further strengthens our conclusion that the ring looks wider in shorter wavelength. 

This experiment gives us confidence that our inferred optical depths and disk geometry remain robust even in the presence of highly reflective grains. Such a conclusion applies to the specific geometry of the 69au ring, but may not be broadly applicable to other geometry.

\section{DustPy Simulation}
\label{sec:dustpy}

\begin{figure*}
    \centering
    \includegraphics[width=0.95\linewidth]{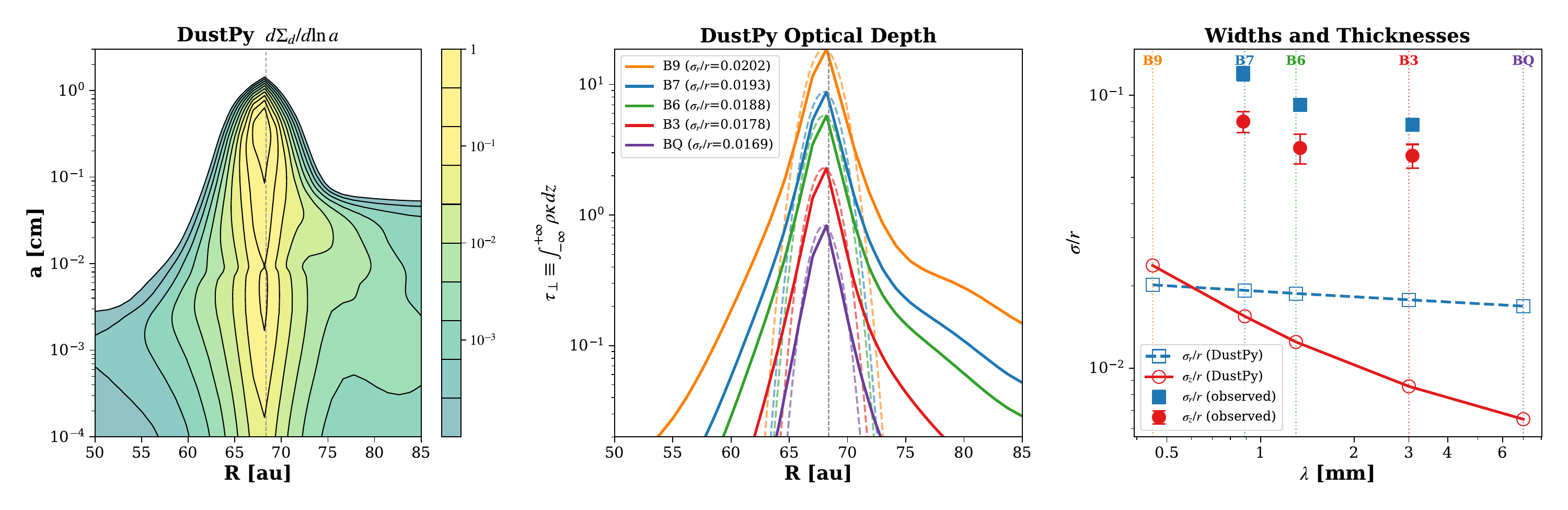}
    \caption{Comparing results of DustPy simulations against data. The left panel shows the simulated dust surface density as a function of grain size and radius, after 1 Myr of evolution. Most of the dust mass has been collected into ring center and are in the form of $\sim 1\cm$ grains. The middle panel shows the resultant (two-sided) optical depths, in various ALMA bands. The right panel exhibits the standard deviations for the DustPy ring (in open symbols) in the radial and vertical directions, as well as our reported values in 3 ALMA bands (Table \ref{tab:fitting}). The former  do decrease with increasing wavelengths, but are much smaller than the observed ones. }
    \label{fig:plot_dustpy2}
\end{figure*}

We carry out DustPy \citep{Stammler2022} simulations to showcase the expectations of dust coagulation models. We adopt a gas density distribution that is centered at 69au, with a scale height $h/r=0.06$, a width $w_r/r = 0.12 \sim 2 h/r$ and a peak gas surface density of $\Sigma = 30 \g/\cm^2$. These are motivated by what we find for the "small grains" (Fig. \ref{fig:fitting}). Dust grains are initialized with the same spatial distribution, at a constant mass fraction of $Z = 0.01$ and at a size of one micron. We further adopt 
a turbulence strength $\alpha = 10^{-4}$, and a fragmentation threshold of $1\m/\s$ \citep[similar to Model 2 of][]{Sierra2025}. Dust grains in our ring quickly conglomerate. Most of the mass has been incorporated into cm-sized grains after a mere 0.2 Myrs. These grains are radially concentrated. Fig. \ref{fig:plot_dustpy2} shows a snapshot of the dust distribution at 1 Myr. 
To compare directly against observation, we assume each size grains are distributed vertically with a scale height $\sigma_h/h = \sqrt{\alpha/St}$, where $St$ is the mid-plane Stokes number.   
We then sum over all sizes (weighted by the simulated mass density) to calculate the 2-D distribution of $\rho_d \kappa_d$, and measure its radial and vertical widths. Fig. \ref{fig:plot_dustpy2} shows our results. According to DustPy, the ring should appear narrow and thin, with fractional widths and heights of order a percent or so. Although these values also decrease with wavelength as the observed ones do, they cannot explain the observed values both in magnitudes and in decay rates.

\end{document}